\documentclass[11pt]{article}
\usepackage[dvips]{graphicx}
\usepackage{enumitem}
\usepackage{amssymb}
\usepackage{color}
\usepackage{amsmath}
\usepackage{nicefrac}
\usepackage[left]{lineno}
\usepackage{hyperref}
\usepackage{adjustbox}
\usepackage{xspace}
\usepackage{multirow}
\usepackage{alphalph}
\usepackage{tabularx}
\usepackage[dvipsnames]{xcolor}

\definecolor{blu}{rgb}{0.,0.,1.}
\definecolor{red}{rgb}{1.,0.,0.}
\definecolor{burgundy}{rgb}{0.5, 0.0, 0.13}
\definecolor{crimsonred}{rgb}{0.6, 0.0, 0.0}
\definecolor{persianblue}{rgb}{0.11, 0.22, 0.73}
\definecolor{forestgreen}{rgb}{0.13,0.35,0.13}

\def\geant {\mbox{\textsc{Geant4}}\xspace}

\hypersetup{colorlinks, citecolor=crimsonred, linkcolor=persianblue, urlcolor=crimsonred}

\begin{document}
\centerline{\LARGE EUROPEAN ORGANIZATION FOR NUCLEAR RESEARCH}

%
\vspace{10mm} {\flushright{
CERN-EP-2023-247 \\
3 November 2023\\
\vspace{4mm}
Revised version:\\15 January 2024\\
}}
\vspace{-30mm}

%

%
\vspace{40mm}

\begin{center}
\boldmath
{\bf {\Large\boldmath{Measurement of the $K^+\to\pi^+\gamma\gamma$ decay}}}
\unboldmath
\end{center}
\vspace{4mm}
\begin{center}
{\Large The NA62 Collaboration}\\
\end{center}

\begin{abstract}
A sample of 3984 candidates of the $K^+\to\pi^+\gamma\gamma$ decay, with an estimated background of $291\pm14$ events, was collected by the NA62 experiment at CERN during 2017--2018. In order to describe the observed di-photon mass spectrum, the next-to-leading order contribution in chiral perturbation theory was found to be necessary.
The decay branching ratio in the full kinematic range is measured to be $(9.61\pm0.17)\times10^{-7}$. The first search for production and prompt decay of an axion-like particle with gluon coupling in the process $K^+\to\pi^+a$, $a\to\gamma\gamma$ is also reported.
\end{abstract}

\begin{center}
{\it Accepted for publication in Physics Letters B}
\end{center}




\newpage

\section*{Introduction}

Experimental studies of radiative non-leptonic kaon decays test chiral perturbation theory (ChPT), which describes low-energy QCD processes. For the $K^+\to\pi^+\gamma\gamma$ decay (denoted $K_{\pi\gamma\gamma}$ below), the ChPT description has been developed at both leading and next-to-leading orders~\cite{thkpigg2,thkpigg3,thkpigg4}.

The $K_{\pi\gamma\gamma}$ decay was first observed by the BNL E787 experiment~\cite{e787}, which reported 31~candidates in the kinematic region 100~MeV$/c < p^*_\pi < 180$~MeV$/c$, where $p^*_\pi $ is the $\pi^+$ momentum in the $K^+$ rest frame. This corresponds to $0.157 < z < 0.384$, where $z=m^2_{\gamma\gamma}/m^2_K$, $m_{\gamma\gamma}$ is the di-photon mass, and $m_K$ is the charged kaon mass. Samples of 149 and 232 $K_{\pi\gamma\gamma}$ decay candidates in the kinematic region $z>0.2$ were selected from the datasets collected at CERN by the NA48/2 experiment in 2003--2004 and the NA62 experiment in 2007, respectively~\cite{na48,na62_2007}. A search near the endpoint in the region $p^*_\pi>213$~MeV$/c$, i.e. $z < 0.048$, was performed by the BNL~E949 experiment~\cite{e949}. All the experimental results reported so far are consistent with the leading-order ChPT description.

A study of the $K_{\pi\gamma\gamma}$ decay based on 3984~candidates in the kinematic region $z>0.2$ collected by the NA62 experiment at CERN in 2017--2018 is reported here. The analysis of the di-photon mass spectrum is performed in the ChPT framework to measure the decay branching ratio and the parameter which characterises the spectrum. The first dedicated search for production and prompt decay of an axion-like particle with gluon coupling in the process $K^+\to\pi^+a$, $a\to\gamma\gamma$ is also reported.


\section{Beam, detector, and data sample}
\label{sec:detector}

\begin{figure}[tb]
\centering
\resizebox{\textwidth}{!}{\includegraphics{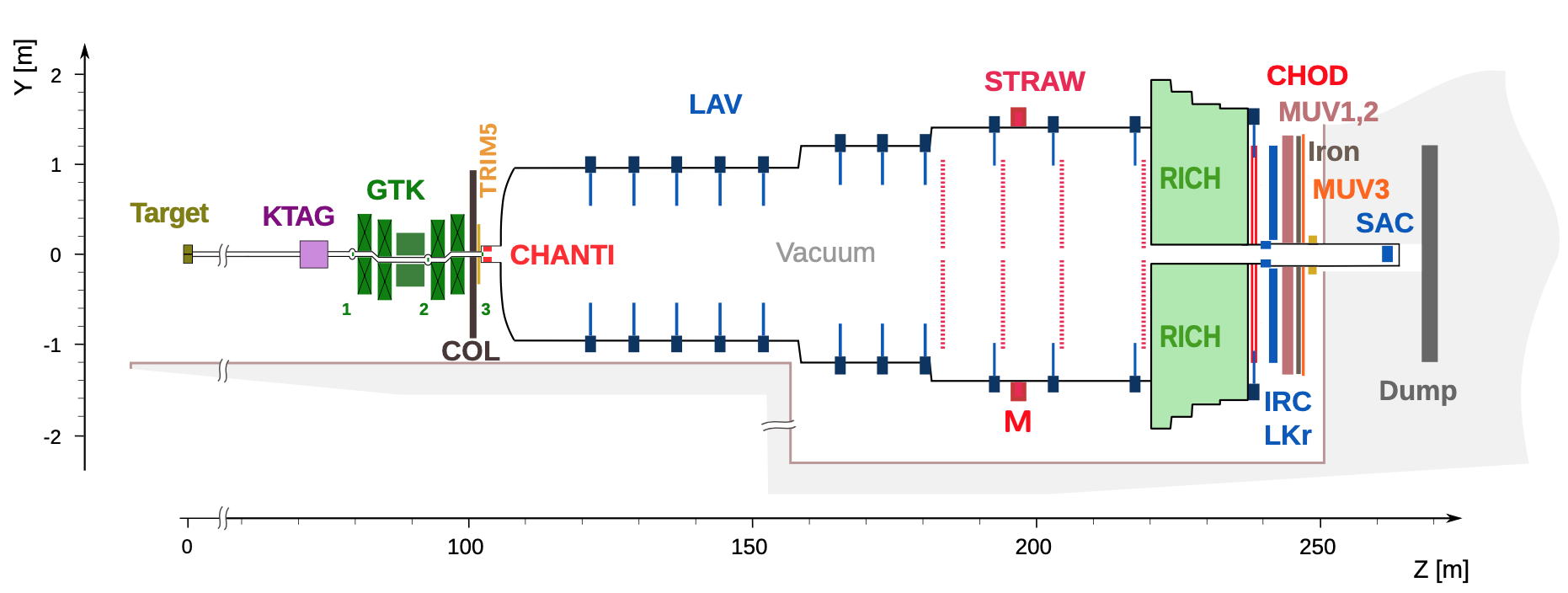}}
\vspace{-10mm}
\caption{Schematic side view of the NA62 beamline and detector used in 2018.}
\label{fig:detector}
\end{figure}

The layout of the NA62 beamline and detector~\cite{na62-detector} is shown schematically in Fig.~\ref{fig:detector}. An unseparated secondary beam of $\pi^+$ (70\%), protons (23\%) and $K^+$ (6\%) is created by directing 400~GeV/$c$ protons extracted from the CERN SPS onto a beryllium target in spills of 4.8~s duration. The central momentum of the beam is 75~GeV/$c$, with a spread of 1\%~(rms).

Beam kaons are tagged with a time resolution of 70~ps by a differential Cherenkov counter (KTAG), which uses as radiator nitrogen gas at 1.75~bar pressure contained in a 5~m long vessel. Beam particle positions, momenta and times (to better than 100~ps resolution) are measured by a silicon pixel spectrometer consisting of three stations (GTK1,2,3) and four dipole magnets forming an achromat. 
A 1.2~m thick steel collimator (COL) with a $76\times40$~mm$^2$ central aperture and $1.7\times1.8$~m$^2$ outer dimensions is placed upstream of GTK3 to absorb hadrons from upstream $K^+$ decays; a variable-aperture collimator of \mbox{$0.15\times0.15$~m$^2$} outer dimensions was used up to early 2018. A dipole magnet (TRIM5) providing a 90~MeV/$c$ horizontal momentum kick is located in front of GTK3. Inelastic interactions of beam particles in GTK3 are detected by an array of scintillator hodoscopes (CHANTI). The beam is delivered into a vacuum tank evacuated to a pressure of $10^{-6}$~mbar, which contains a 75~m long fiducial volume (FV) starting 2.6~m downstream of GTK3. The beam angular spread at the FV entrance is 0.11~mrad (rms) in both horizontal and vertical planes. Downstream of the FV, undecayed beam particles continue their path in vacuum.


The momenta of charged particles produced in $K^+$ decays in the FV are measured by a magnetic spectrometer (STRAW) located in the vacuum tank downstream of the FV. The spectrometer consists of four tracking chambers made of straw tubes, and a dipole magnet (M) located between the second and third chambers that provides a horizontal momentum kick of 270~MeV/$c$ in a direction opposite to that produced by TRIM5. The momentum resolution is $\sigma_p/p = (0.30\oplus 0.005\cdot p)\%$, with the momentum $p$ expressed in GeV/$c$.

\newpage

%
A ring-imaging Cherenkov detector (RICH) consisting of a 17.5~m long vessel filled with neon at atmospheric pressure (with a Cherenkov threshold of 12.5~GeV/$c$ for pions) provides particle identification, charged particle time measurements with a typical resolution of 70~ps, and the trigger time.
Two scintillator hodoscopes (CHOD), which include a matrix of tiles and two planes of slabs arranged in four quadrants located downstream of the RICH, provide trigger signals and time measurements with 200~ps precision.

A 27$X_0$ thick quasi-homogeneous liquid-krypton (LKr) electromagnetic calorimeter is used for particle identification and photon detection. The calorimeter has an active volume of 7~m$^3$, segmented in the transverse direction into 13248 projective cells of
$2\times 2$~cm$^2$ size, and provides an energy resolution $\sigma_E/E=(4.8/\sqrt{E}\oplus11/E\oplus0.9)\%$, with $E$ expressed in GeV. To achieve hermetic acceptance for photons emitted in $K^+$ decays in the FV at angles up to 50~mrad from the beam axis, the LKr calorimeter is complemented by annular lead glass detectors (LAV) installed in 12~positions inside and downstream of the vacuum tank, and two lead/scintillator sampling calorimeters (IRC, SAC) located close to the beam axis. An iron/scintillator sampling hadronic calorimeter formed of two modules (MUV1,2) and a muon detector consisting of 148~scintillator tiles located behind an 80~cm thick iron wall (MUV3) are used for particle identification.

%
%

The data sample analysed is obtained from $8.3\times 10^5$ SPS spills recorded in 2017--2018, with the typical beam intensity increasing over time from \mbox{$1.8\times 10^{12}$} to \mbox{$2.2\times 10^{12}$} protons per spill. The latter value corresponds to a 500~MHz mean instantaneous beam particle rate at the FV entrance, and a 3.7~MHz mean $K^+$ decay rate in the FV.

The main NA62 trigger line is designed for the measurement of the ultra-rare $K^+\to\pi^+\nu\bar\nu$ decay~\cite{pinn}. The present analysis is based on dedicated control and non-muon trigger lines operating concurrently with the main trigger line and downscaled typically by factors of 400 and 200, respectively. A multi-track (MT) trigger line downscaled typically by a factor of 100 is used for evaluation of systematic uncertainties. The trigger description is provided in Refs.~\cite{am19,na62-trigger}. The control line requires only CHOD signals. The non-muon and MT lines both include  low-level hardware (L0) and high-level software (L1) triggers. For the non-muon line, the L0 condition includes RICH signal multiplicity in coincidence with a CHOD signal, in the absence of in-time MUV3 signals. For the MT line, the L0 condition includes RICH signal multiplicity and coincidence of signals in two diagonally opposite CHOD quadrants. The L1 algorithm for both lines involves beam $K^+$ identification by the KTAG and reconstruction of a STRAW track originating in the FV. 

Monte Carlo (MC) simulations of particle interactions with the detector and its response are performed using a software package based on the \geant toolkit~\cite{geant4}. Accidental activity and trigger responses are included in the simulation.


\section{Event selection}
\label{sec:selection}
\vspace{-0.8mm}

The abundant $K^+\to\pi^+\pi^0$ decay followed by the prompt $\pi^0\to\gamma\gamma$ decay (denoted $K_{2\pi}$), collected concurrently with signal candidates through the same trigger lines, is used for normalisation. Signal and normalisation decay modes have the same set of particles in the final state, leading to a first-order cancellation of detector and trigger inefficiencies, and reducing the systematic uncertainties in the measurement. The principal selection criteria are listed below.
\vspace{-0.4mm}
\begin{itemize}
\item A positively charged track reconstructed with the STRAW spectrometer is considered as a $\pi^+$ candidate, if it has the following properties: the track time, evaluated using the RICH or CHOD signals spatially associated with the track, should be within 2~ns of the trigger time; the track momentum is required to be in the range 15--65 GeV/$c$; the track’s trajectory through the STRAW chambers and its extrapolation to the LKr calorimeter, RICH and CHOD should be within the geometrical acceptances of these detectors. Events with more than one $\pi^+$ candidate are rejected.
\vspace{-0.2mm}
\item The parent kaon is defined by a KTAG signal with time~($t_\text{KTAG}$) within 1~ns of the $\pi^+$ time, and a reconstructed GTK track with time~($t_\text{GTK}$) within 0.5~ns of the KTAG signal. Association between the GTK and STRAW tracks relies on a discriminant built using the time difference $\Delta t=t_\text{GTK}-t_\text{KTAG}$ and the closest distance of approach (CDA) of the pion and kaon tracks. The GTK track with the discriminant most consistent with a $K^+ - \pi^+$ decay vertex
is identified as the parent kaon. The reconstructed kaon decay vertex, defined as the point of closest approach of the GTK and STRAW tracks, is required to be within the FV.
\vspace{-0.2mm}
\item Events are rejected if they have additional STRAW tracks compatible with originating from the reconstructed kaon decay vertex.
%
\vspace{-0.2mm}
\item Particle identification conditions are applied to the $\pi^+$ candidate as follows: the ratio of the energy deposited in the LKr calorimeter, $E$, to the momentum, $p$, measured by the STRAW spectrometer should be $E/p < 0.85$; signals in the MUV3 detector geometrically associated with the $\pi^+$ candidate are not allowed within 4~ns of $t_\text{KTAG}$.
\vspace{-0.2mm}
\item Photon candidates are defined as clusters of energy deposited in the LKr calorimeter within 5~ns of the $\pi^+$ candidate time with energy exceeding 2~GeV, not associated with any STRAW track, and separated by at least 250~mm from the $\pi^+$ track impact point on the LKr calorimeter front plane. Exactly two photon candidates are required. The distance between the candidates should exceed 250~mm in the LKr calorimeter front plane. Photon four-momenta are computed using cluster energies, cluster positions and the kaon decay vertex position.
\vspace{-0.2mm}
\item Multi-photon backgrounds with merged clusters in the LKr calorimeter are suppressed by a requirement on the photon candidate cluster size. The cluster size is defined as the rms lateral width along the axis in the LKr front plane corresponding to the maximum width value. Cluster sizes of both photon candidates should be less than 20~mm.
\vspace{-0.2mm}
\item The transverse momentum of the $\pi^+\gamma\gamma$ system with respect to the GTK track direction should be $p_T<30$~MeV$/c$.
\vspace{-0.2mm}
\item Events with $0.2<z<0.51$ ($0.04<z<0.12$) are identified as $K_{\pi\gamma\gamma}$ ($K_{2\pi}$) candidates. The kinematic variable $z$ is computed using the track information as $z=(P_K-P_\pi)^2/m_K^2\equiv m^2_{\gamma\gamma}/m^2_K$, where $P_K$ and and $P_\pi$ are the reconstructed $K^+$ and $\pi^+$ four-momenta. The photon candidate information is not used, which reduces systematic uncertainties and improves the $z$ resolution; the  resolution varies  from $\delta z=3.5\times 10^{-3}$ at $z=0.2$ to zero at the endpoint $z=0.515$.
\item The reconstructed mass, $m_{\pi\gamma\gamma}$, and the difference between the reconstructed momentum of the $\pi^+\gamma\gamma$ system and the reconstructed parent kaon momentum, $\Delta p=p_{\pi\gamma\gamma}-p_K$, are required to be inside an ellipse centred on the point ($m_K$, 0) with semi-axes of 10~MeV/$c^2$ and 3~GeV/$c$, respectively. The axes of the ellipse are oriented to account for the correlation between $m_{\pi\gamma\gamma}$ and $\Delta p$. This condition has an efficiency of 97.3\% for both $K_{\pi\gamma\gamma}$ and $K_{2\pi}$.
\end{itemize}

\begin{figure}[p]
\centering
\includegraphics[width=0.47\textwidth]{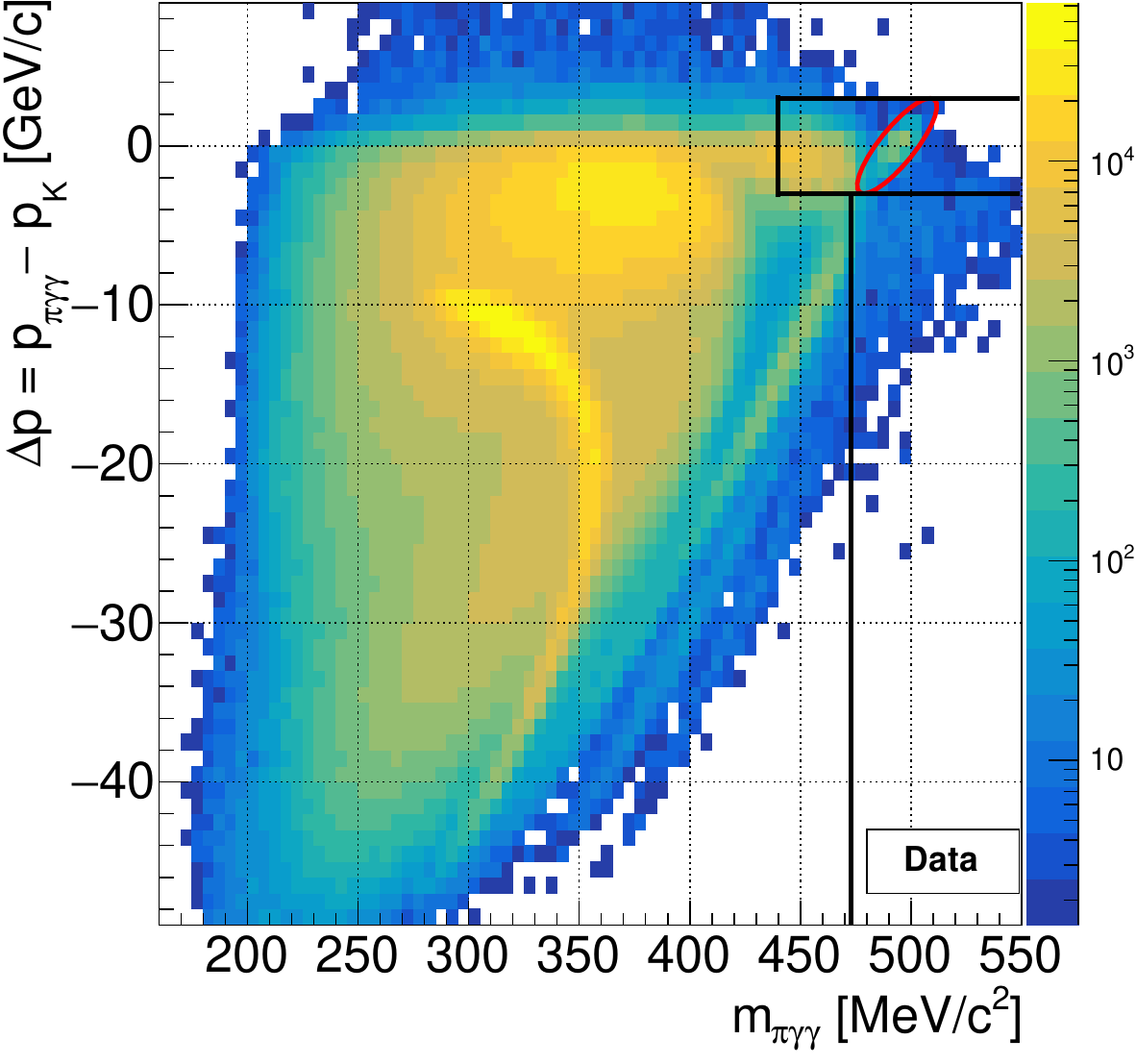}%
\includegraphics[width=0.47\textwidth]{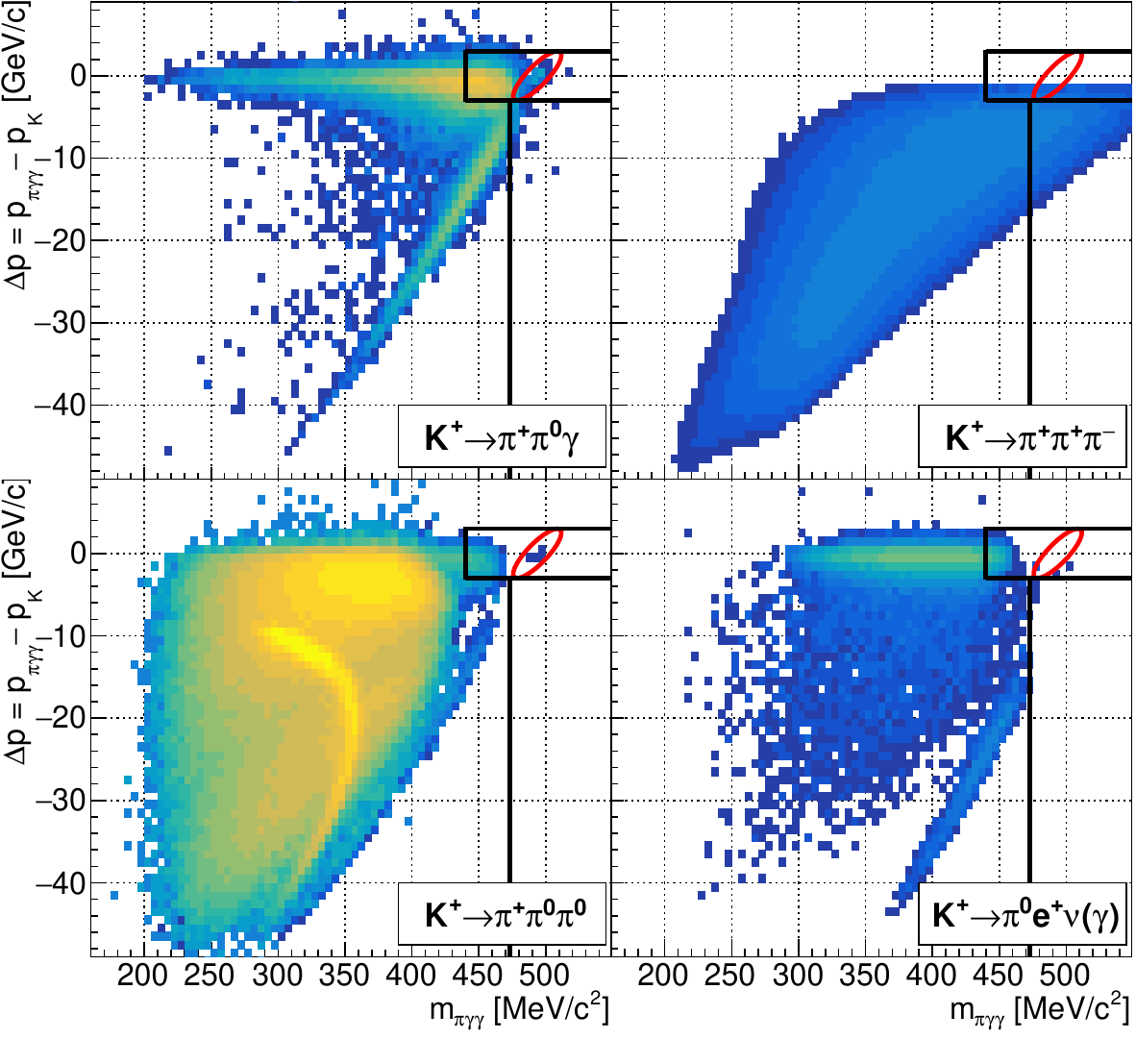}
\vspace{-5mm}
\caption{Reconstructed ($m_{\pi\gamma\gamma}$, $\Delta p$) distributions after the full $K_{\pi\gamma\gamma}$ selection
for data (left) and simulated background decay samples (right). The simulated distributions have a common normalisation to the data according to their branching ratios; all plots have the same colour scale. The elliptical selection condition, and the rectangular regions used to obtain the $m_{\pi\gamma\gamma}$ spectrum 
(Fig.~\ref{fig:mass_plots}) and a control data sample for $K^+\to\pi^+\pi^+\pi^-$ background evaluation~(as discussed in Section~\ref{sec:background}) are shown by solid lines.}
\label{fig:dp_dm}
\end{figure}

\begin{figure}[p]
\centering
\includegraphics[width=0.5\textwidth]{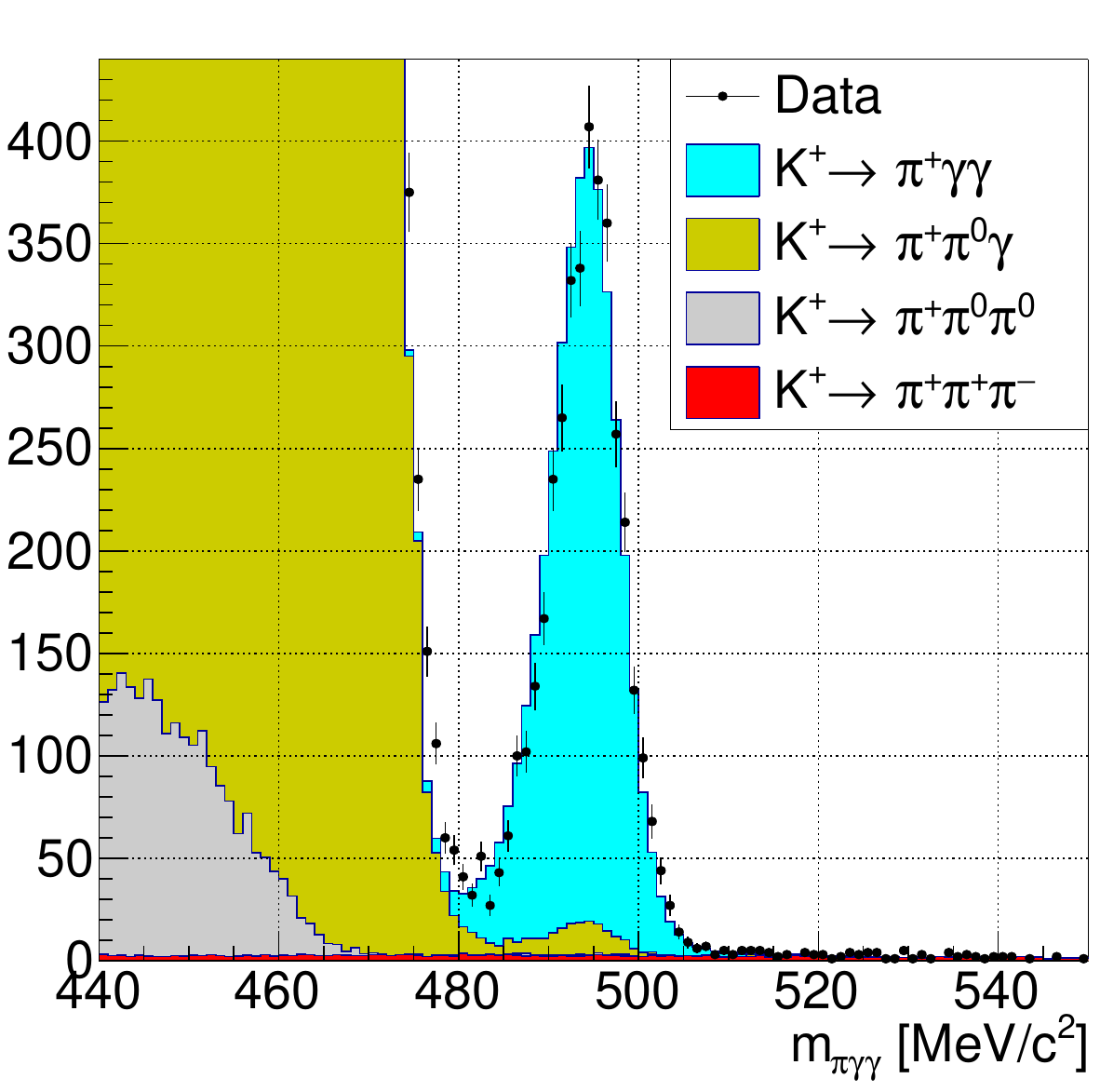}%
\includegraphics[width=0.5\textwidth]{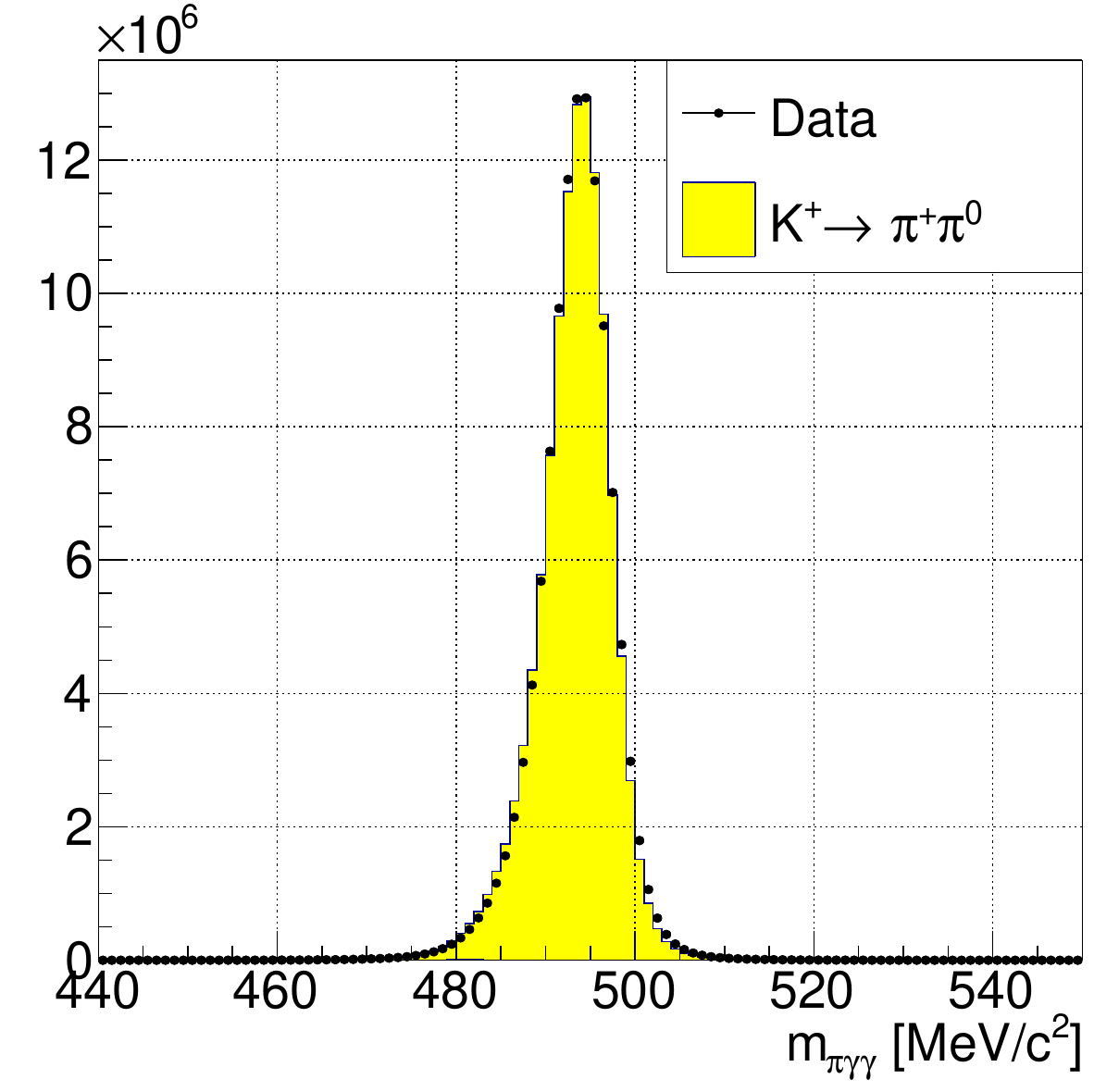}
\vspace{-10mm}
\caption{Reconstructed $m_{\pi\gamma\gamma}$ mass distributions of the $K_{\pi\gamma\gamma}$~(left) and $K_{2\pi}$~(right) candidates for selected samples of data and simulated signal and background samples. The elliptical selection condition is replaced by the requirement $|\Delta p|<3$~GeV/$c$. The simulated $K_{\pi\gamma\gamma}$ signal corresponds to the result of the ChPT fit.}
\label{fig:mass_plots}
\end{figure}

The reconstructed ($m_{\pi\gamma\gamma}$, $\Delta p$) distributions of the $K_{\pi\gamma\gamma}$ candidates for data and for simulated signal and background samples are displayed in Fig.~\ref{fig:dp_dm}: 3984 candidates are observed in the signal region in the data. The reconstructed $m_{\pi\gamma\gamma}$ distributions of the selected $K_{\pi\gamma\gamma}$ and $K_{2\pi}$ candidates, obtained using a modified selection with the elliptical condition in the ($m_{\pi\gamma\gamma}$, $\Delta p$) replaced by the requirements $|\Delta p|<3$~GeV/$c$, $440<m_{\pi\gamma\gamma}<550$~MeV$/c^2$, are shown in Fig.~\ref{fig:mass_plots}.


\section{Trigger efficiencies and number of kaon decays}

The trigger lines used in the analysis, the fractions of the dataset collected with each line, and the trigger efficiencies measured for the data using control datasets and modelled for the simulated samples are summarised in Table~\ref{tab:trigger}. Part of the dataset was collected with the STRAW condition disabled in the non-muon trigger line. The $K_{\pi\gamma\gamma}$ and $K_{2\pi}$ decays have different $\pi^+$ momentum spectra. 
The efficiency of the control trigger has no significant geometric or momentum dependence; therefore it cancels between signal and normalisation samples and is not simulated.
In contrast,
the efficiency of the non-muon line is momentum dependent due to the RICH and STRAW conditions and differs between the two decay modes. The response of the non-muon trigger is modelled for simulated samples, and is included in the acceptance.

\begin{table}[p]
\centering
\vspace{-3mm}
\caption{Trigger lines, fractions of the dataset collected with each line, and trigger efficiencies for the $K_{\pi\gamma\gamma}$ and $K_{2\pi}$ samples, $\varepsilon(K_{\pi\gamma\gamma})$ and $\varepsilon(K_{2\pi})$. Statistical uncertainties are quoted for the $K_{\pi\gamma\gamma}$  trigger efficiencies in data, and are negligible for other quantities.}
\vspace{1mm}
\begin{tabular}{l|c|cc|cc}
\hline
Trigger line &
Fraction &
\multicolumn{2}{c}{Data} &
\multicolumn{2}{c}{Simulation} \\
\cline{3-6}
& of dataset & 
$\varepsilon(K_{\pi\gamma\gamma})$ &
$\varepsilon(K_{2\pi})$ &
$\varepsilon(K_{\pi\gamma\gamma})$ &
$\varepsilon(K_{2\pi})$ \\
\hline
Control & 39\% & $(99.1\pm0.2)\%$ & 99.1\% & -- & -- \\
Non-muon & 36\% & $(96.7\pm0.6)\%$ & 98.0\% & 96.5\% & 97.9\% \\
Non-muon {\small (no L1-STRAW)} & 25\% & $(99.5\pm0.2)\%$ & 99.5\% & 99.1\% & 99.6\% \\
\hline
\end{tabular}
\vspace{-2mm}
\label{tab:trigger}
\end{table}


The effective number of $K^+$ decays in the FV is calculated as
\begin{displaymath}
N_K = \frac{N_{K2\pi}\cdot D_{K2\pi}}{A_{K2\pi}\cdot\mathcal{B}(K_{2\pi})\cdot\mathcal{B}(\pi^0_{\gamma\gamma})}=(5.55\pm0.03)\times 10^{10},
\end{displaymath}
where $N_{K2\pi}=1.14\times 10^8$ is the number of data events passing the $K_{2\pi}$ selection; $D_{K2\pi}=10$ is a software downscaling factor applied to the $K_{2\pi}$ candidates in the data at the processing stage;
$A_{K2\pi}=(10.05\pm0.05)\%$ is the acceptance of the $K_{2\pi}$ selection evaluated with simulations, with a systematic uncertainty evaluated by variation of the selection conditions; $\mathcal{B}(K_{2\pi})=(20.67\pm0.08)\%$ and $\mathcal{B}(\pi^0_{\gamma\gamma})=(98.823\pm0.034)$\% are the branching ratios in the $K_{2\pi}$ decay chain~\cite{pdg}. The relative background in the $K_{2\pi}$ sample is found with simulations to be of ${\cal O}(10^{-5})$ and is neglected. 



\begin{figure}[t]
\centering
\includegraphics[width=0.71\textwidth]{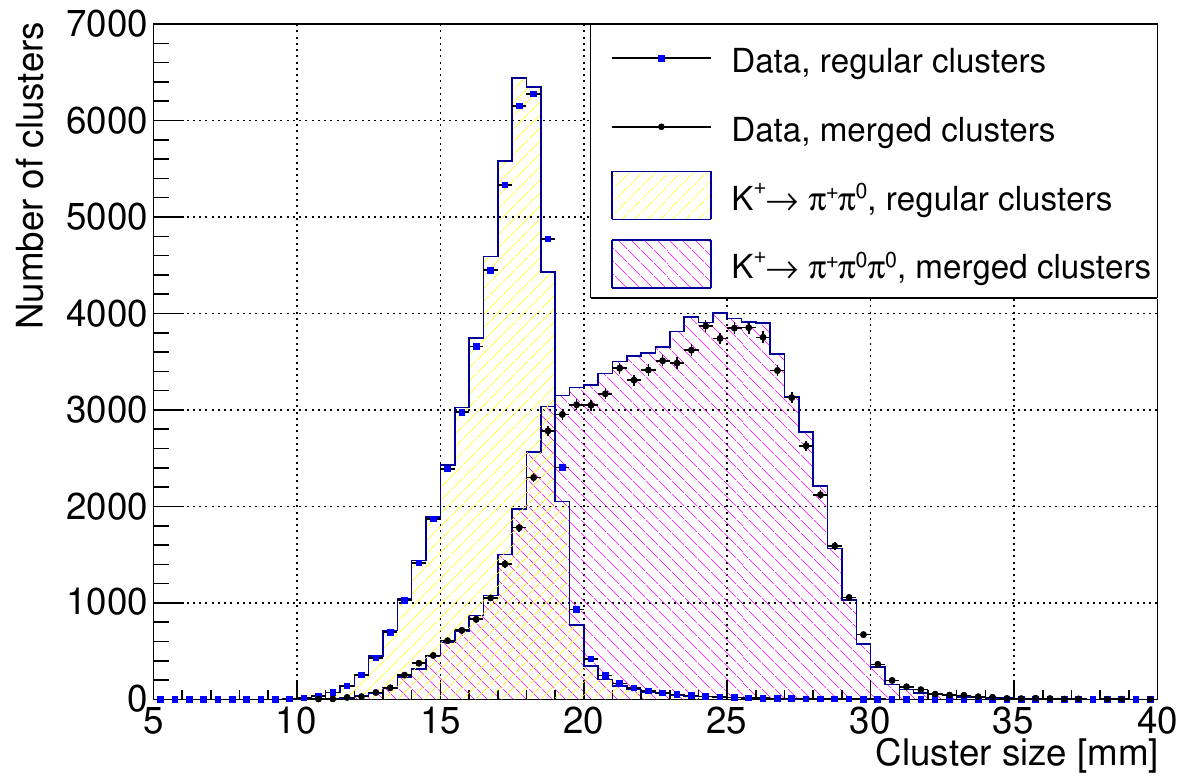}
\vspace{-5mm}
\caption{Cluster size distributions for regular and merged LKr clusters for data and simulated samples. An arbitrary scale factor of $2\times10^{-4}$ is applied to the regular cluster distributions to fit the scale. The accuracy of the simulation is limited both for regular and merged clusters.}
\label{fig:clustersize}
\end{figure}


\boldmath
\section{Backgrounds to the $K_{\pi\gamma\gamma}$ decay}
\unboldmath
\label{sec:background}

\subsection*{Multi-photon backgrounds}

Multi-photon backgrounds come from $K^+\to\pi^+\pi^0\gamma$~($K_{2\pi\gamma}$) and
$K^+\to\pi^+\pi^0\pi^0$~($K_{3\pi0}$) decays followed by $\pi^0\to\gamma\gamma$ decays, where the photons produce overlapping showers in the LKr calorimeter leading to the  reconstruction of merged clusters. In the $K_{2\pi\gamma}$ case, a merged cluster is formed by the photon from the $K^+$ decay and a photon from the $\pi^0$ decay. In the $K_{3\pi0}$ case, two merged clusters are formed. These conditions lead to final states with no missing energy, and the background $m_{\pi\gamma\gamma}$ distribution peaks at the $K^+$ mass as for signal events (Fig.~\ref{fig:dp_dm}, right). 

To validate the simulation of cluster merging, a modified $K_{\pi\gamma\gamma}$ selection is employed. Three photon candidates are required, the cluster size conditions are removed, and the reconstructed mass of each photon pair 
is required to differ from the $\pi^0$ mass by more than 8~MeV$/c^2$. The selected data sample consists mainly of $K_{3\pi0}$ events with a single merged cluster.
The merged cluster (M) is identified as that with reconstructed energy lying within 1~GeV of its expected value based on the energies and positions of the other two clusters~(A, B):
\begin{displaymath}
E_{\rm M}=m^2_{\pi0}\,L^2\cdot(1/(E_{\rm A}d^2_{\rm AM})+1/(E_{\rm B}d^2_{\rm BM}))\,.
\end{displaymath}
Here $E_{\rm A(B)}$ are the reconstructed energies of the clusters A(B), $d_{\rm A(B)M}$ are the distances between the clusters A(B) and M, and $L$ is the distance between the $K^+$ decay vertex and the LKr calorimeter front plane.

Cluster size distributions for regular (non-merged) clusters from $K_{2\pi}$ decays and merged clusters from $K_{3\pi0}$ decays for data and simulated samples are displayed in Fig.~\ref{fig:clustersize}. The cluster size requirement, smaller than 20~mm, 
removes 2.9\% (2.5\%) of the regular clusters for data (simulated) events. The number of merged clusters identified in the data is 80519, while $84103\pm1200$ are expected from a simulated $K_{3\pi0}$ sample normalised to the number of kaon decays ($N_K$) evaluated using the $K_{2\pi}$ sample. With the cluster size requirement applied, 18820 merged clusters remain in the data, compared to $19953\pm 281$ expected from simulation. This validates the modelling of cluster merging to a 5\% precision. The uncertainties in the expected values are dominated by the systematic uncertainty in the $K_{3\pi0}$ branching ratio. A systematic uncertainty of 5\% in relative terms is assigned to the estimated $K_{2\pi\gamma}$ and $K_{3\pi0}$ backgrounds.


\boldmath
\subsection*{Background from the  $K^+\to\pi^+\pi^+\pi^-$ decay}
\unboldmath

The $K^+\to\pi^+\pi^+\pi^-$ ($K_{3\pi}$) decays contribute to the background when two pions ($\pi^+$ and $\pi^-$) are not reconstructed by the STRAW spectrometer while producing clusters in the LKr calorimeter. To satisfy the $p_{\pi\gamma\gamma}$ selection condition, the non-reconstructed pions must deposit most of their energy in the calorimeter. This background contributes in the kinematic region $z>0.3$ above the di-pion threshold.

Hadronic showers in the LKr calorimeter are imperfectly modelled by the \geant toolkit. The $K_{3\pi}$ background is therefore simulated using a calorimeter response model obtained from the data. The probability for a $\pi^\pm$ to create exactly one cluster in the LKr calorimeter with a size smaller than 20~mm is measured as a function of both the $\pi^\pm$ momentum and the energy deposited in the calorimeter using a sample of  $K_{3\pi}$ decays selected kinematically from the STRAW spectrometer information.

\noindent To validate the $K_{3\pi}$ background simulation two control samples are used:
\begin{itemize}

\item Sample A consists of $K_{3\pi}$ decays where two pions deposit most of their energy in the LKr calorimeter with negligible background. This sample is obtained from a modified $K_{\pi\gamma\gamma}$ selection in which each photon candidate is required to have an associated STRAW track, and the three STRAW tracks in the event are required to be kinematically compatible with a $K_{3\pi}$ decay. To increase the sample size, the MT trigger line is employed in addition to the control and non-muon trigger lines used for the main analysis.


\item Sample B consists mainly of $K_{3\pi}$ events with two pions not reconstructed by the STRAW spectrometer and depositing only a fraction of their energy in the LKr calorimeter. This sample is obtained from a modified $K_{\pi\gamma\gamma}$ selection where the elliptical condition in the ($m_{\pi\gamma\gamma}$, $\Delta p$) plane is replaced by the requirements $m_{\pi\gamma\gamma}-m_K>-20$~MeV/$c^2$ and $\Delta p<-3$~GeV/$c$ (as shown in Fig.~\ref{fig:dp_dm}), and additionally requiring $z>0.3$. These conditions are introduced to suppress the $K_{3\pi0}$ and $K_{2\pi\gamma}$ contributions.
\end{itemize}

The two control samples do not overlap with the $K_{\pi\gamma\gamma}$ signal sample. The numbers of events observed in the control samples in the data, and those expected from simulations of the $K_{3\pi}$, $K_{2\pi\gamma}$ and $K^+\to\pi^0e^+\nu(\gamma)$ decays, are presented in Table~\ref{tab:bkg2}. Simulations are in agreement with the data, which validates the $K_{3\pi}$ background model within a 3\% statistical precision determined by the size of the control data samples. A systematic uncertainty of 3\% in relative terms is assigned to the estimated $K_{3\pi}$ background.


\begin{table}[b]
\centering
\vspace{-6mm}
\caption{Numbers of events expected from simulations and observed in the data in the control samples A and B used for validation of the $K_{3\pi}$ background evaluation.}
\vspace{1mm}
\begin{tabular}{lr@{}c@{}rr@{}c@{}r}
\hline
Source & \multicolumn{3}{c}{Sample A} & \multicolumn{3}{c}{Sample B} \\
\hline
$K^+\to\pi^+\pi^+\pi^-$ & 933 & $\pm$ & $9_\text{stat}$  & 1007 & $\pm$ &$5_\text{stat}$ \\
$K^+\to\pi^+\pi^0\gamma$ & & -- & & 33 & $\pm$ &$8_\text{stat}$ \\
$K^+\to\pi^0 e^+\nu(\gamma)$ & & -- & & 23 & $\pm$ &$2_\text{stat}$ \\
$K^+\to\pi^+\pi^0\pi^0$ & & -- & & 1 & $\pm$ &$1_\text{stat}$ \\
\hline
Total expected & 933 & $\pm$ & $9_\text{stat}$  & 1064 & $\pm$ &$10_\text{stat}$ \\ \hline
Data & & 917 & & & 1044 & \\
\hline
\end{tabular}
\vspace{-3mm}
\label{tab:bkg2}
\end{table}

\begin{table}[b!]
\centering
\caption{Estimated background contributions in the $K_{\pi\gamma\gamma}$ sample.}
\vspace{1mm}
\begin{tabular}{lrcrcr}
\hline
Source & \multicolumn{5}{c}{Estimated background} \\
\hline
$K^+\to\pi^+\pi^0\gamma$ & 240& $\pm$ & $8_\text{stat}$ &$\pm$ & $12_\text{syst}$ \\
$K^+\to\pi^+\pi^+\pi^-$ & 35 &$\pm$ & $1_\text{stat}$ & $\pm$ & $1_\text{syst}$ \\
$K^+\to\pi^+\pi^0\pi^0$ & 9 & $\pm$ & $2_\text{stat}$ & & \\
$K^+\to \pi^0 e^+\nu(\gamma)$ & 7 &$\pm$ & $1_\text{stat}$ & & \\
\hline
Total & 291 &$\pm$ & $8_\text{stat}$ & $\pm$ & $12_\text{syst}$ \\
\hline
\end{tabular}
\label{tab:signalregion}
\end{table}


\subsection*{Other backgrounds and summary}

The $K^+\to\pi^0 e^+\nu(\gamma)$ decay contributes to the background when the $e^+$ is misidentified as a $\pi^+$, one of the photons from the $\pi^0\to\gamma\gamma$ decay is not detected, and either a radiative photon from the $K^+$ decay or a photon produced by bremsstrahlung is detected. Background from the $K^+\to\pi^+e^+e^-$ decay where the $e^\pm$ tracks are not reconstructed but produce LKr clusters is found to be negligible.

In total, 3984 $K_{\pi\gamma\gamma}$ candidates are observed in the data. The background contributions in the signal sample estimated with simulations are reported in Table~\ref{tab:signalregion}. 
The estimated number of background events is $291\pm 8_{\rm stat}\pm 12_{\rm syst}$, which corresponds to 7\% of the signal candidates.


\boldmath
\section{Measurement of the $K_{\pi\gamma\gamma}$ decay}
\unboldmath
\label{sec:chpt}

The $K_{\pi\gamma\gamma}$ decay is described by the two kinematic variables
\begin{equation*}
y=\frac{P_K(P_{\gamma1}-P_{\gamma2})}{m^2_K}, \quad z=\frac{m^2_{\gamma\gamma}}{m^2_K},
\end{equation*}
where $P_K$ and $P_{\gamma1,2}$ are the four-momenta of the kaon and the two photons, $m_{\gamma\gamma}$ is the di-photon mass and $m_K$ is the $K^+$ mass. The physical region of the kinematic variables is
\begin{equation*}
|y|\leq\frac{1}{2}\lambda^{1/2}(1,r^2_\pi,z), \quad 0\leq z\leq(1-r_\pi)^2=0.515,
\end{equation*}
where $\lambda(a,b,c)=a^2+b^2+c^2-2(ab+ac+bc)$, $r_\pi=m_\pi/m_K$, and $m_\pi$ is the $\pi^+$ mass.
In the ChPT framework at next-to-leading order, $\mathcal{O}(p^6)$, the decay rate is 
parameterised as follows~\cite{thkpigg3}:
\begin{displaymath}
\frac{\partial^2\Gamma}{\partial y\partial z}=\frac{m_K}{2^9\pi^3}\left[z^2\left(|A(\hat{c},z,y^2)+B(z)|^2+|C(z)|^2\right)+\left(y^2-\frac{1}{4}\lambda(1,r^2_\pi,z)\right)^2|B(z)|^2\right],
\end{displaymath}
where $A(\hat{c},z,y^2)$ and $B(z)$ are the loop amplitudes (the latter dominates at low $z$ and vanishes at leading order), and $C(z)$ is the pole amplitude contributing a few percent to the total decay rate. The differential decay rate depends strongly on the $z$ variable, and only weakly on the $y$ variable. The parameter $\hat{c}$ is the only free parameter in this analysis. The decay amplitudes depend on external parameters, namely the effective weak octet coupling $G_8$ fixed in this analysis according to~\cite{g8}, the $K_{3\pi}$ decay amplitude parameters ($\alpha_1$, $\alpha_3$, $\beta_1$, $\beta_3$, $\gamma_3$, $\zeta_1$, $\xi_1$) fixed according to~\cite{k3pipar}, and the polynomial contributions $\eta_i$ fixed to zero as in the previous measurements. The values of the external parameters are summarised in Table~\ref{tab:extparam}.


\begin{table}[!b]
\centering
\vspace{-5mm}
\caption{Values of the external parameters used for this and previous $K_{\pi\gamma\gamma}$ measurements. The parameter values are reported in~\cite{thkpigg2,g8,k3pipar,Kambor:1991ah,Bijnens:2002vr}.}
\vspace{1mm}
\begin{tabular}{lrrr}
\hline
Parameter & E787~\cite{e787} & NA48/2, NA62~\cite{na48,na62_2007} & This measurement \\
\hline
$G_8m^2_K\times10^6$ & 2.24 & 2.202 & 2.202 \\
$\alpha_1\times10^8$ & 91.71 & 93.16 & 92.80 \\
$\alpha_3\times10^8$ & $-7.36$ & $-7.62$ & $-7.45$ \\
$\beta_1\times10^8$ & $-25.68$ & $-27.06$ & $-26.46$ \\
$\beta_3\times10^8$ & $-2.43$ & $-2.22$ & $-2.50$ \\
$\gamma_3\times10^8$ & 2.26 & 2.95 & 2.78 \\
$\zeta_1\times10^8$ & $-0.47$ & $-0.40$ & $-0.11$ \\
$\xi_1\times10^8$ & $-1.51$ & $-1.83$ & $-1.20$ \\
$\eta_i$~($i=1$, 2, 3) & 0 & 0 & 0 \\
\hline
\end{tabular}
\vspace{-1mm}
\label{tab:extparam}
\end{table}

\begin{figure}[t]
\centering
\includegraphics[width=0.5\textwidth]{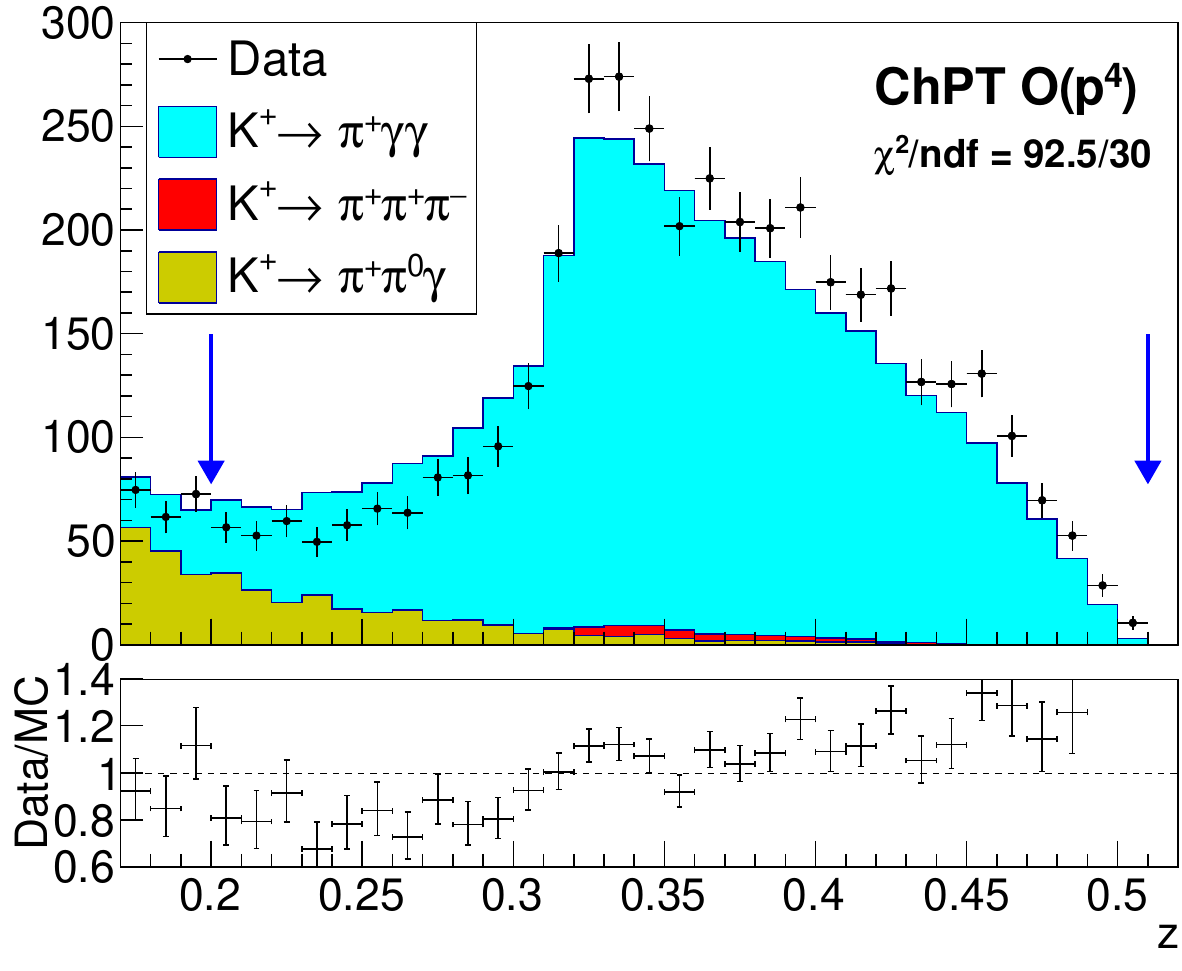}%
\includegraphics[width=0.5\textwidth]{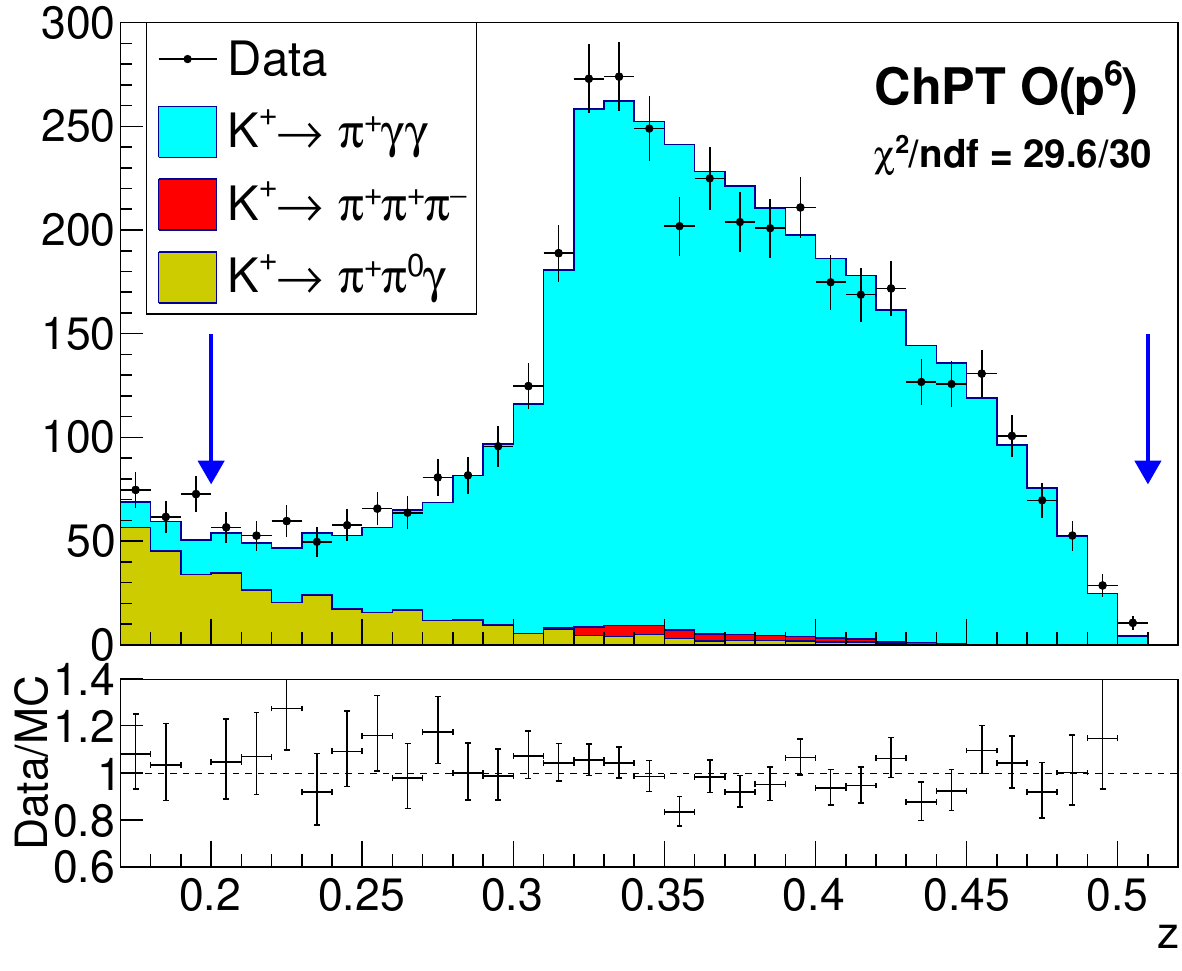}
\vspace{-9mm}
\caption{Reconstructed $z$ spectrum of the $K_{\pi\gamma\gamma}$ candidates, estimated background contributions  and simulated signal spectra using the $\hat{c}$ values obtained from the fits (upper panels). 
The signal region used in the fit is indicated with arrows.
  The two panels correspond to the $\mathcal{O}(p^4)$ and $\mathcal{O}(p^6)$ signal descriptions. 
  The ratios of data and simulated spectra with their full uncertainties are shown in the lower panels.}
\label{fig:fit}
\end{figure}

To measure the parameter $\hat{c}$, the reconstructed $z$ spectrum of the signal candidates (Fig.~\ref{fig:fit}) is fitted according to the prescription of~\cite{thkpigg3} in 31~bins of equal width in the signal region $0.2<z<0.51$ to find the minimum of the quantity
\begin{equation*}
\chi^2 = (\vec{k}-\vec{\lambda}(\hat{c}))^T\cdot C^{-1}\cdot (\vec{k}-\vec{\lambda}(\hat{c})),
\label{eq:chi2}
\end{equation*}
where the vector $\vec{k}$ contains the numbers of observed data events in the $z$ bins, and the vector $\vec{\lambda} = \vec\lambda_S(\hat{c})+ \vec\lambda_B$ contains the simulated numbers of the signal, $\vec\lambda_S(\hat{c})$, and background, $\vec\lambda_B$, events in the $z$ bins for a given $\hat{c}$ value. The covariance matrix $C$ accounts for the statistical uncertainties in bins of the data and simulated $z$ distributions, the systematic uncertainty due to $N_K$ which is fully correlated among the bins, and the systematic uncertainties in the multi-photon and $K_{3\pi}$ background estimates are also conservatively assumed to be fully correlated.
Additional systematic uncertainties in $\hat c$ may arise from: the measurement of the trigger efficiency; the LKr calorimeter energy calibration; the resolution of the $z$ variable.
Each of these is found to be $\delta\hat c<10^{-3}$ and neglected.

Fits to the data using both leading order, $\mathcal{O}(p^4)$, and next-to-leading order, $\mathcal{O}(p^6)$, ChPT descriptions~\cite{thkpigg3} are performed. The resulting fit $p$-values are $2.7\times10^{-8}$ and 0.49, respectively. This constitutes the first evidence that the $\mathcal{O}(p^4)$ description is not compatible with the data.
Fit results corresponding to the minimum $\chi^2$ values are shown in Fig.~\ref{fig:fit}. 

Model-independent $K_{\pi\gamma\gamma}$ partial branching ratios, $\mathcal{B}_i$, and differential decay widths, $(d\Gamma/dz)_i$, 
are computed in the 31~bins of the $z$ variable in the signal region as
\begin{displaymath}
\mathcal{B}_i=\frac{N_i-N^{\rm B}_i}{N_K\cdot A_i}, \quad\quad
\left(\frac{d\Gamma}{dz} \right)_i=\frac{\Gamma_K\mathcal{B}_i}{\Delta z},
\end{displaymath}
where $N_i$ is the number of $K_{\pi\gamma\gamma}$ candidates observed in the $i$-th bin in data, $N^{\rm B}_i$ the estimated number of background events in that bin, $A_i$ the signal acceptance in that bin,
$N_K$ the number of $K^+$ decays in the FV, $\Gamma_K=(5.32\pm0.01)\times 10^{-17}$~GeV
the charged kaon decay width~\cite{pdg}, and $\Delta z=0.01$ is the bin width. The statistical errors in $A_i$, the effects of the dependence of $A_i$ on the assumed $K_{\pi\gamma\gamma}$ kinematic distribution and the resolution effects are negligible with respect to the statistical uncertainties in $N_i$.


\begin{table}[p]
\centering
\caption{Systematic uncertainties for $\hat{c}$, $\mathcal{B}$ and $\mathcal{B}_\text{MI}(z>0.2)$ measurements.}
\vspace{1mm}
\begin{tabular}{lccc}
\hline
Source & $\delta\hat{c}$ & $\delta\mathcal{B}\times 10^7$ & $\delta\mathcal{B}_\text{MI}(z>0.2)\times 10^7$ \\
\hline
Number of kaon decays & 0.026 & 0.056 & 0.064 \\
Simulation of multi-photon backgrounds & 0.016& 0.034 & 0.026 \\
Simulation of $K_{3\pi}$ background & 0.001 & 0.002 & 0.003 \\
Limited size of simulated samples & 0.014 & 0.030 & 0.018 \\
\hline
Total & 0.034 & 0.072 & 0.072 \\
\hline
\end{tabular}
\label{tab:syst}
\end{table}


\begin{figure}[p]
\centering
\vspace{-1mm}
\includegraphics[width=0.6\textwidth]{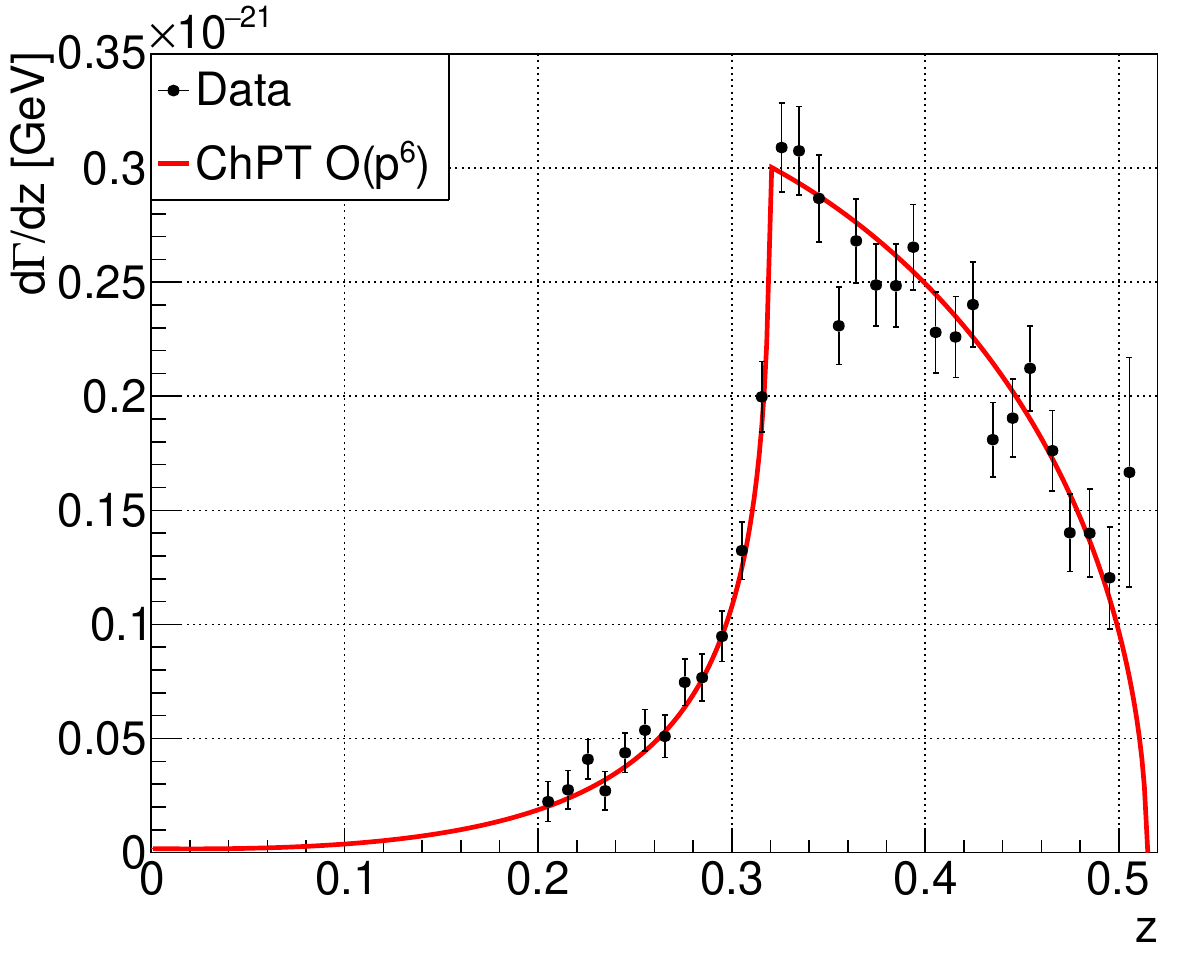}
\vspace{-6mm}
\caption{Differential $K_{\pi\gamma\gamma}$ decay width as a function of $z$. Markers: model-independent measurements in the signal region with their full uncertainties. The $z$ positions of the markers are evaluated according to~\cite{xcoord}. Solid line: differential decay width in the ChPT $\mathcal{O}(p^6)$ description with $\hat{c}=1.144$. External parameters of the ChPT description are fixed as shown in Table~\ref{tab:extparam}.}
\label{fig:mivsc6}
\end{figure}

\begin{figure}[p]
\centering
\includegraphics[width=0.5\textwidth]{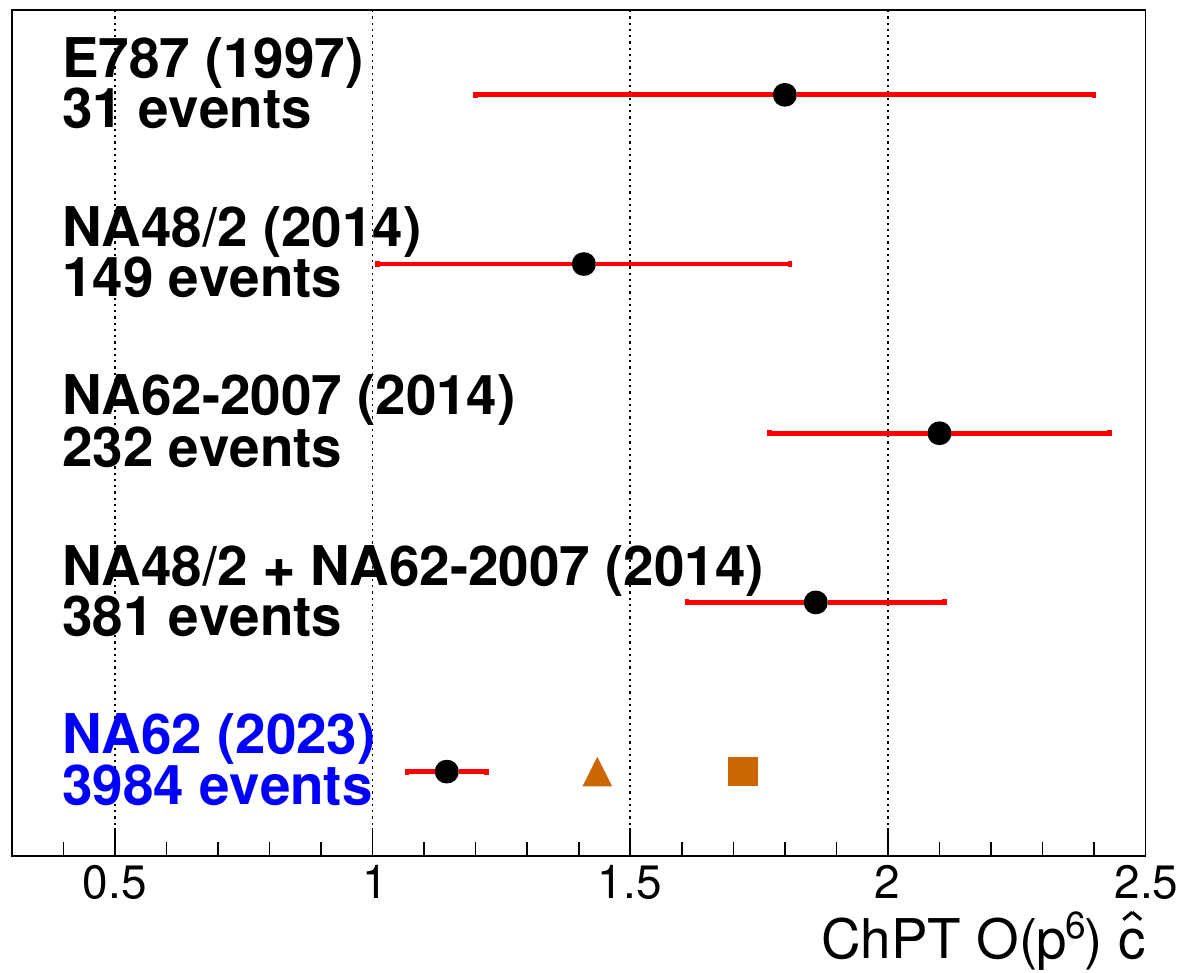}%
\includegraphics[width=0.5\textwidth]{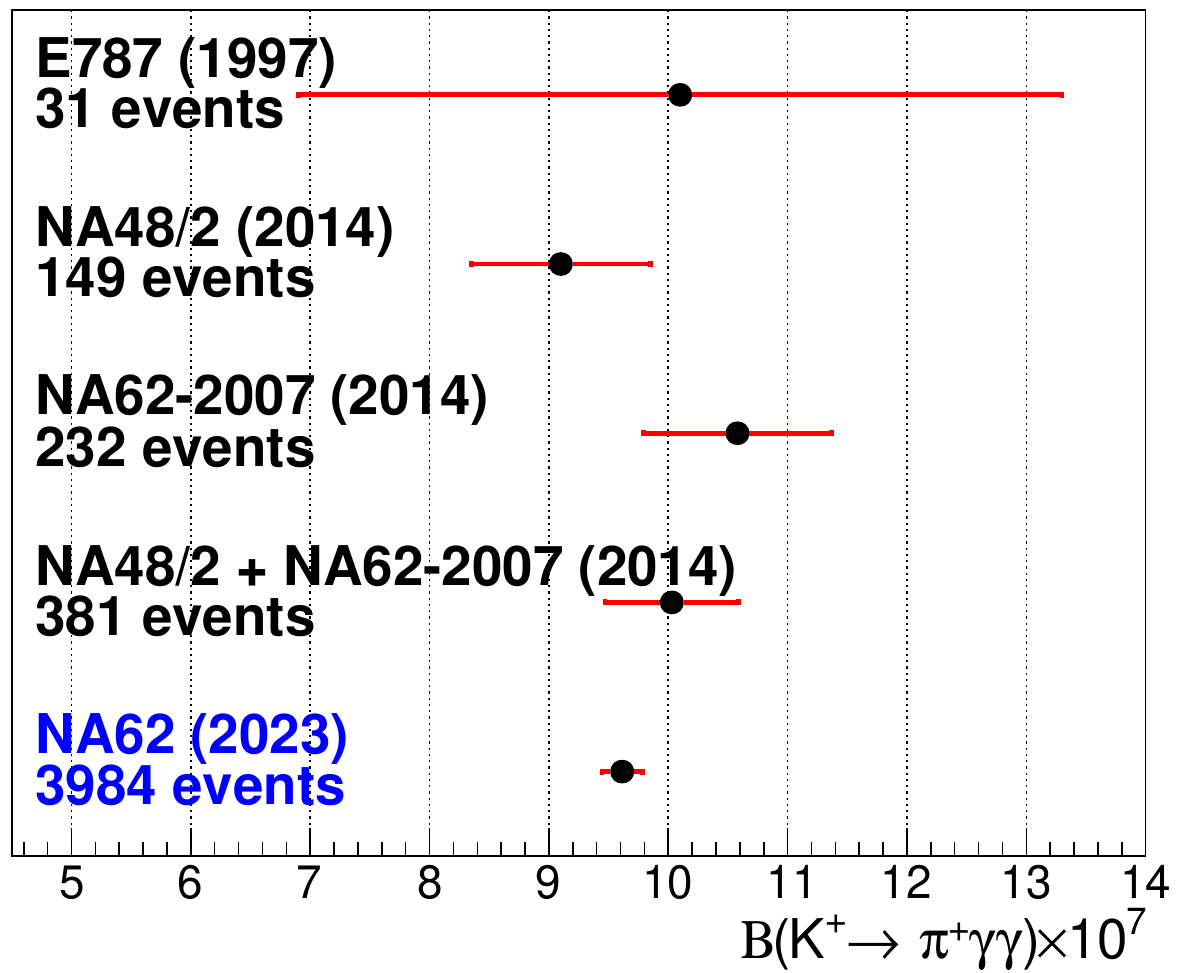}
\vspace{-8mm}
\caption{Summary of the $\hat{c}$~(left) and $\mathcal{B}$~(right) measurements in the ChPT $\mathcal{O}(p^6)$ framework. The error bars do not include uncertainties due to the external parameters of the ChPT fit. The present measurement, earlier measurements~\cite{e787,na48,na62_2007} and the average of the two measurements~\cite{na48,na62_2007} reported in~\cite{na62_2007} are shown. The triangle~(square) shows the central $\hat{c}$ value obtained in the present analysis considering external parameter values used by E787~\cite{e787} (NA48/2 and NA62-2007~\cite{na48,na62_2007}), respectively.}
\label{fig:finalresults}
\end{figure}


\boldmath
\section{Results of the $K_{\pi\gamma\gamma}$ measurement}
\unboldmath
\label{sec:results}

A sample of 3984 $K_{\pi\gamma\gamma}$ decay candidates with an estimated background contamination of $291\pm14$ events is analysed. The observed $z$ spectrum is consistent with the ChPT description to the next-to-leading order, $\mathcal{O}(p^6)$. The $\hat{c}$ parameter is measured in the ChPT $\mathcal{O}(p^6)$ description to be
\begin{displaymath}
\hat{c} = 1.144\pm0.069_\text{stat}\pm0.034_\text{syst}.
\end{displaymath}
The corresponding branching ratio obtained by integration of the ChPT $\mathcal{O}(p^6)$ differential branching ratio in the full kinematic range is
\begin{displaymath}
\mathcal{B} = (9.61\pm0.15_\text{stat}\pm0.07_\text{syst})\times10^{-7}.
\end{displaymath}
The model-independent branching ratio in the region $0.2<z<0.51$ is measured by summing over the $z$ bins to be
\begin{displaymath}
\mathcal{B}_\text{MI}(z>0.2)=(9.46\pm0.19_\text{stat}\pm0.07_\text{syst})\times10^{-7}.
\label{eq:finalbrmi}
\end{displaymath}
The statistical uncertainties are due to the limited size of the signal sample. The individual contributions to the systematic uncertainties are listed in Table~\ref{tab:syst}. The $d\Gamma/dz$ spectrum corresponding to the central value of $\hat c$ obtained from the ChPT fit, and the differential decay widths in the $z$ bins obtained from the model-independent measurement, are displayed in Fig.~\ref{fig:mivsc6}.

The ChPT $\mathcal{O}(p^6)$ decay amplitudes $A(\hat{c},z,y^2)$ and $B(z)$ depend on external parameters, which are fixed in this analysis as summarised in Table~\ref{tab:extparam}. The quantities $\hat{c}$ and $\mathcal{B}$ measured in the ChPT $\mathcal{O}(p^6)$ framework are compared to the previous measurements~\cite{e787,na48,na62_2007} in Fig.~\ref{fig:finalresults}. Also shown are the $\hat c$ values obtained using the same sets of external parameter values as each of the previous measurements (Table~\ref{tab:extparam}): the $\hat c$ value depends significantly on the external parameter values. The dependence arises mainly from the sensitivity to~$\xi_1$, with the derivative $\partial\hat c/\partial\xi_1=-1.0\times 10^8$. On the other hand, the sensitivity of $\mathcal{B}$ to external parameter values is negligible with respect to the uncertainties quoted above, the largest derivative being $\partial{\cal B}/\partial\zeta_1=-1.8$. Uncertainties in $\hat c$ and $\mathcal{B}$ due to those in the external parameters are not quoted.

The measured values of $\hat c$, $\cal B$ and ${\cal B}_{\rm MI}(z>0.2)$ are consistent within one standard deviation with the earlier measurements~\cite{e787,na48,na62_2007} when the same sets of external parameter values are used, and are obtained with an improved precision.


\vspace{-0.5mm}
\boldmath
\section{Search for the $K^+\to\pi^+a$, $a\to\gamma\gamma$ decay}
\unboldmath
\label{sec:kpix}

An axion-like particle (ALP, $a$) coupling to gluons is considered in one of the hidden-sector scenarios known as BC11~\cite{Beacham:2019nyx,fips1,fips2}. For ALP masses $m_a<3m_\pi$, the ALP decays almost exclusively into photon pairs. A search for the $K^+\to\pi^+a$, $a\to\gamma\gamma$ decay chain is performed as a peak search in the distribution of the missing mass of the $K_{\pi\gamma\gamma}$ candidates, $m_\text{miss}=\sqrt{(P_K-P_\pi)^2}$, where $P_K$ and $P_\pi$ are the reconstructed $K^+$ and $\pi^+$ four-momenta. 

The signal is investigated in 287~mass hypotheses in the search range of 207--350~MeV$/c^2$, with a step of 0.5~MeV$/c^2$. The lower limit of the range is chosen to avoid background from the non-gaussian tail of the $K_{2\pi}$ decay. The signal region in each $m_a$ hypothesis is defined as $|m_\text{miss}-m_a|<1.5\sigma_m$, where $\sigma_m$ is 
the $m_\text{miss}$ resolution evaluated with simulations. The resolution improves from 2.0~MeV/$c^2$ to 0.2~MeV/$c^2$ across the search range.

\begin{figure}[t]
\centering
\includegraphics[width=0.5\textwidth]{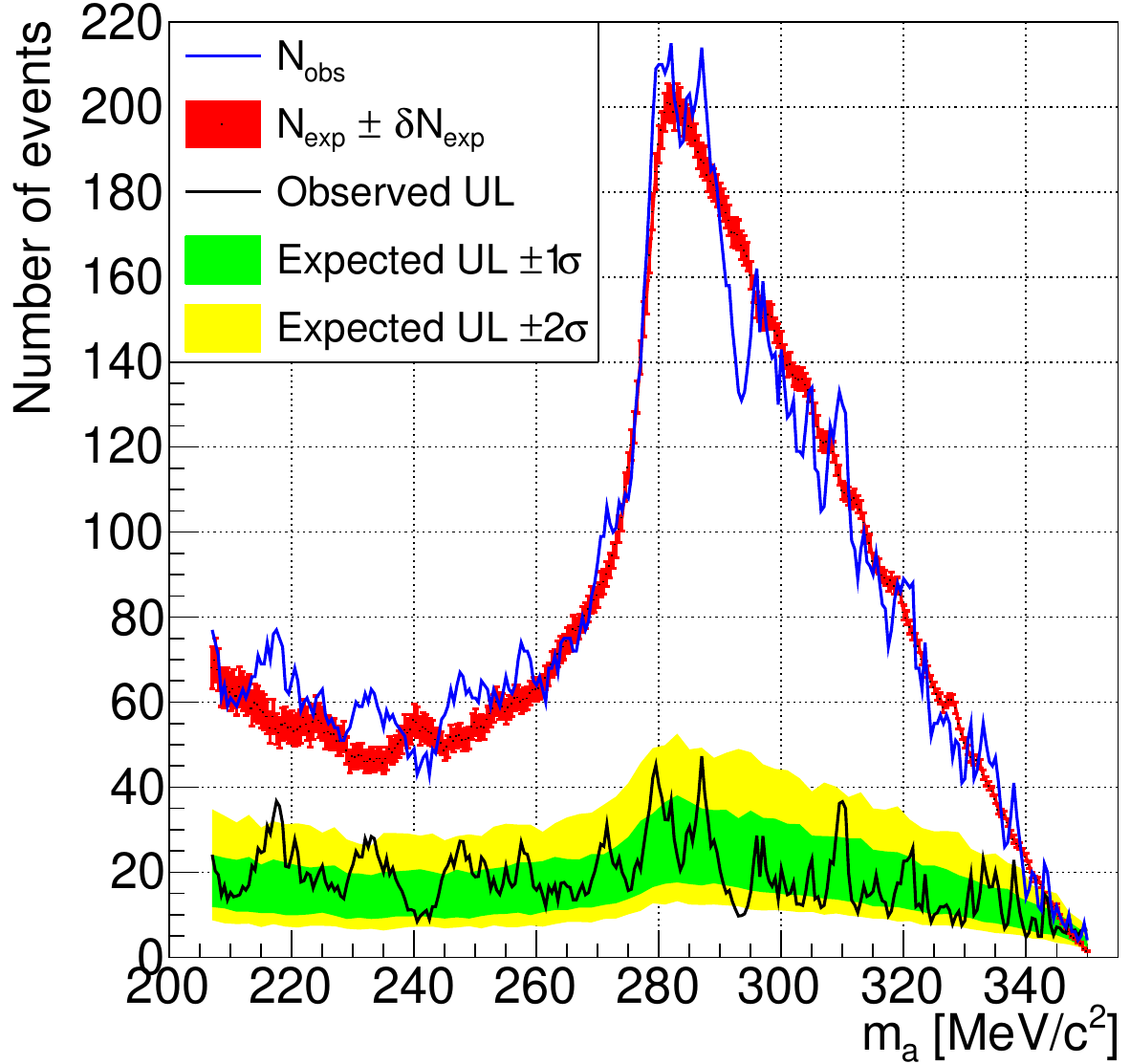}%
\includegraphics[width=0.5\textwidth]{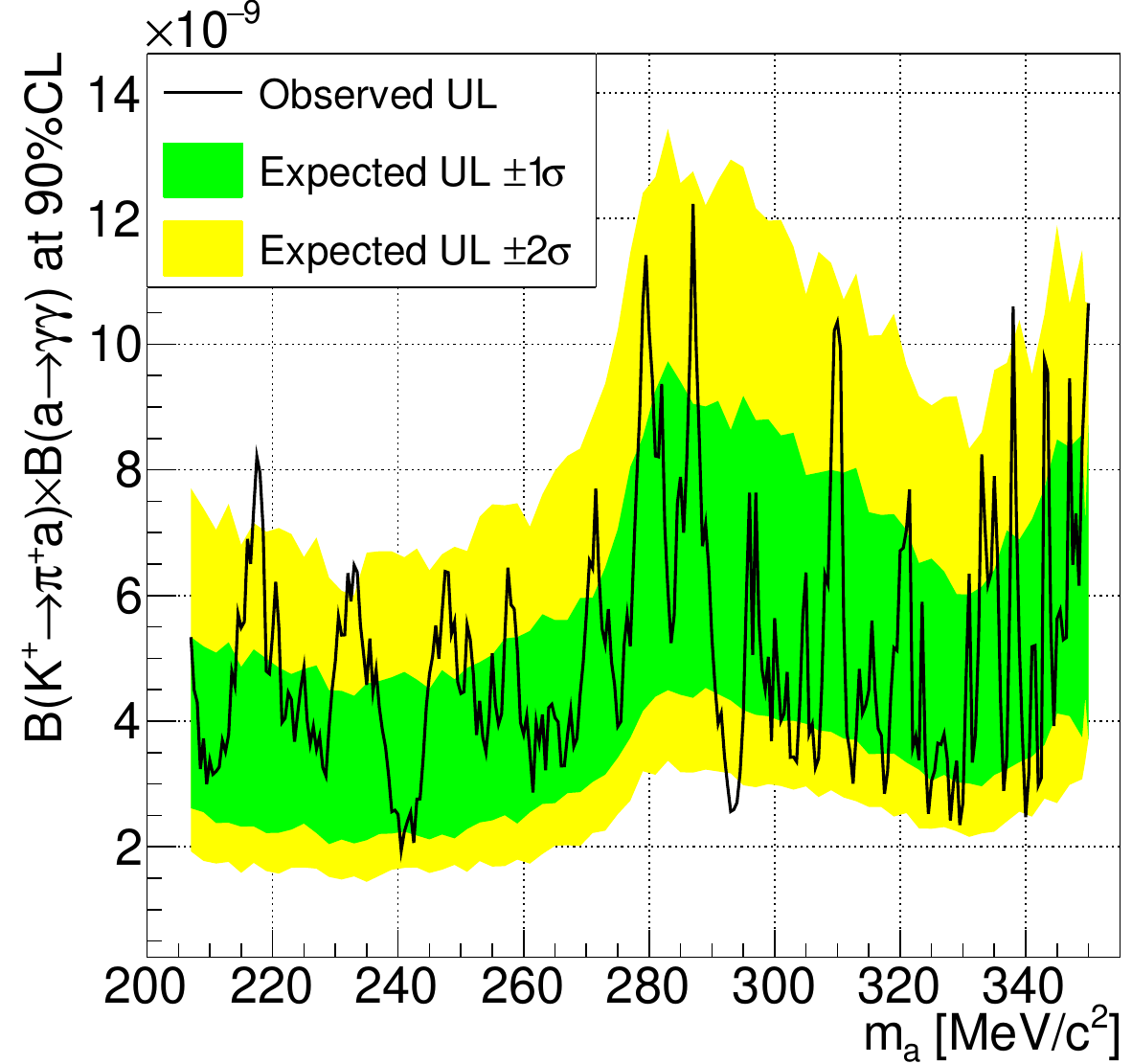}
\vspace{-8mm}
\caption{Search for ALP production and decay. Left: observed and expected numbers of events, and upper limits at 90\% CL of the number of signal events with their expected $\pm1\sigma$ and $\pm2\sigma$ bands in the null hypothesis for each ALP mass value considered. Right: upper limits at 90\%~CL of $\mathcal{B}(K^+\to\pi^+a)\times\mathcal{B}(a\to\gamma\gamma)$ in the prompt decay assumption with their expected bands.}
\label{fig:ulkpix}
\end{figure}

In each mass hypothesis, the  background in the signal region, $N_{\rm exp}$, is determined from simulations including the $K_{\pi\gamma\gamma}$ contribution according to the ChPT $\mathcal{O}(p^6)$ description (Section~\ref{sec:results}) and the backgrounds to the $K_{\pi\gamma\gamma}$ decay (Section~\ref{sec:background}). The uncertainty in the estimated background, $\delta N_{\rm exp}$, is evaluated considering the uncertainties in $\hat c$ and in the contributions from backgrounds to the $K_{\pi\gamma\gamma}$ decay. Upper limits at 90\% CL of the numbers of signal events, $N_S$, are computed using $N_\text{obs}$, $N_\text{exp}$ and $\delta N_\text{exp}$ using the CL$_\text{S}$ method~\cite{cls}. The values of $N_\text{obs}$, $N_\text{exp}$ with $\delta N_\text{exp}$, the observed upper limits of $N_S$, and the expected $\pm1\sigma$ and $\pm2\sigma$ bands of variation of $N_S$ in the null hypothesis are displayed in Fig.~\ref{fig:ulkpix}~(left). 
The sensitivity of the search is limited by the $K_{\pi\gamma\gamma}$ background. No statistically significant evidence for the $K^+\to\pi^+a$, $a\to\gamma\gamma$ decay chain is observed. 


Under the assumption of a prompt $a\to\gamma\gamma$ decay, upper limits of the branching ratio $\mathcal{B}(K^+\to\pi^+a)$ are evaluated in each $m_a$ hypothesis using the relation
\begin{displaymath}
\mathcal{B}(K^+\to\pi^+a) =
\frac{N_S}{N_K\cdot A_S},
\end{displaymath}
where $A_S$ is the selection acceptance for the prompt $a\to\gamma\gamma$ decay evaluated with simulations, which decreases from 8\% to 1\% across the search range. The resulting upper limits at 90\% CL are shown in Fig.~\ref{fig:ulkpix}~(right).


Under the assumption of non-zero ALP mean lifetime, $\tau_a$, the signal selection acceptance decreases as a function of $\tau_a$
due to
underestimation of the quantity $m_{\pi\gamma\gamma}$ for events with displaced $a\to\gamma\gamma$ decay vertices. The signal selection has non-zero acceptance for $a\to\gamma\gamma$ decay vertices displaced up to 10~m with respect to the $K^+$ decay vertex. The $m_{\rm miss}$ resolution does not depend on $\tau_a$ because $m_{\rm miss}$ is evaluated using the GTK and STRAW information only. To account for the reduction of signal acceptance, the upper limits of $\mathcal{B}(K^+\to\pi^+a)$ obtained in the prompt decay assumption are divided by a signal acceptance loss function parameterised empirically as follows:
\begin{displaymath}
f(\tau_a)=\frac{2}{\pi}\arctan\left(\frac{0.1366}{\tau_a}-\frac{0.0042}{\tau^2_a}+\frac{0.0002}{\tau^3_a}\right),
\end{displaymath}
with $\tau_a$ expressed in~ns. This function tends to unity for $\tau_a\to 0$, and decreases with $\tau_a$: $f(0.25)=0.29$, $f(0.5)=0.16$, $f(1)=0.08$, $f(3)=0.03$.
The parameterisation is accurate to a 20\% precision
for $\tau_a<3$~ns for all $m_a$ values within the search range. The search has no sensitivity for $\tau_a>3$~ns as the ALP becomes practically invisible.

\begin{figure}[t]
\centering
\includegraphics[width=0.575\textwidth]{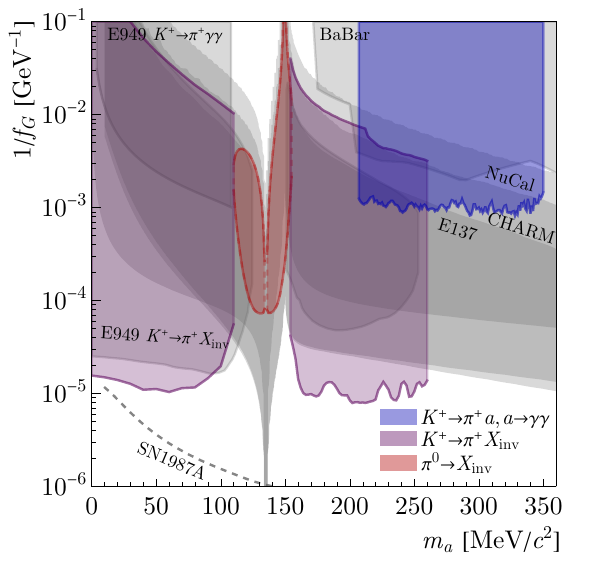}
\put(-79,69){\scriptsize\color{Black}{\textbf{\textsf{NA62 results}}}}
\vspace{-5mm}
\caption{Excluded regions at 90\% CL in the BC11 scenario parameter space $(m_a, 1/f_G)$. Limits obtained from the present search for the $K^+\to\pi^+a$, $a\to\gamma\gamma$ decay, and from earlier NA62 searches for the $K^+\to\pi^+X_{\rm inv}$~\cite{pinn} and $\pi^0\to X_{\rm inv}$~\cite{pi0inv} decays are shown as coloured areas. Other experimental limits on the BC11 scenario~\cite{Afik:2023mhj} are shown as grey shaded areas.}
\vspace{-1mm}
\label{fig:bc11}
\end{figure}

The free parameters of the BC11 scenario are the ALP mass and coupling strength ($m_a$, $1/f_G$). The upper limits obtained at 90\% CL for $\mathcal{B}(K^+\to\pi^+a)$ are interpreted in the BC11 framework considering the dependence of $\mathcal{B}(K^+\to\pi^+a)$ and $\tau_a$ on the free parameters ($m_a$, $1/f_G$) provided in~\cite{bc11_1,bc11_2}. The results are shown in Fig.~\ref{fig:bc11}. The ALP proper mean lifetime scales as $\tau_a\sim f_G^2$, which makes the ALP effectively invisible for $1/f_G<1~{\rm TeV}^{-1}$. 
The search is limited by background, therefore the lower bound of the excluded region scales as $1/f_G\sim N_K^{-1/4}$.

Dedicated searches for the $K^+\to\pi^+X_{\rm inv}$ decay with an invisible particle in the final state performed by the NA62 experiment in the mass range 0--260~MeV/$c^2$~\cite{pinn,pi0inv} also have sensitivity to the long-lived ALP in the BC11 scenario.
The corresponding exclusion regions at 90\% CL on the ALP coupling in the BC11 scenario, obtained using the published upper limits of ${\cal B}(K^+\to\pi^+X_{\rm inv})$~\cite{pinn,pi0inv} accounting for the dependence of the signal acceptance on the assumed ALP lifetime, are shown in Fig.~\ref{fig:bc11}.
The $K^+\to\pi^+X_\text{inv}$ search~\cite{pinn} is not limited by background, is sensitive to a wide $\tau_a$ range, and uses a non-downscaled trigger line. The combined effect of these three factors is a far greater sensitivity to $1/f_G$ than the $K^+\to\pi^+a$, $a\to\gamma\gamma$ search.
There is no sensitivity for $m_a=m_{\pi 0}$ as the ALP decays promptly in this case, and the asymmetry of the exclusion region in the vicinity of $m_{\pi 0}$ is due to the asymmetric behaviour of the ALP lifetime~\cite{bc11_2}. The results obtained are in agreement with an independent interpretation of the NA62 data (Fig.~5-left of~\cite{Afik:2023mhj}). Note that the convention for the ALP coupling used in this analysis and in~\cite{bc11_1,bc11_2,Afik:2023mhj} differs by a factor of $4\pi^2$ from the convention used in~\cite{Beacham:2019nyx,fips1,fips2}.


\section*{Summary}
\vspace{-0.8mm}

A study of the $K^+\to\pi^+\gamma\gamma$ decay is reported by the NA62 experiment at CERN, based on 3984~candidates collected  in 2017--2018 in the kinematic range $z>0.2$ with an estimated background of $291\pm14$ events.  The ChPT contribution at next-to-leading order, ${\cal O}(p^6)$, must be taken into account to describe the observed di-photon mass spectrum. Using this description, the $\hat c$ parameter is measured to be $1.144\pm0.077$, and the decay branching ratio in the full kinematic range is found to be $(9.61\pm0.17)\times10^{-7}$. The first search for production and prompt decay of an axion-like particle with gluon coupling in the process $K^+\to\pi^+a$, $a\to\gamma\gamma$ is also reported. The results are used to establish an exclusion region in the parameter space of the BC11 scenario.


\section*{Acknowledgements}
\vspace{-0.8mm}

We are grateful to Giancarlo D'Ambrosio, Marc Knecht and Siavash Neshatpour for the discussion of the theoretical aspects of the ChPT fits, and to Kohsaku Tobioka for the discussion of ALP production phenomenology in the BC11 scenario and for providing numerical data.
It is a pleasure to express our appreciation to the staff of the CERN laboratory and the technical
staff of the participating laboratories and universities for their efforts in the operation of the
experiment and data processing.

The cost of the experiment and its auxiliary systems was supported by the funding agencies of 
the Collaboration Institutes. We are particularly indebted to: 
F.R.S.-FNRS (Fonds de la Recherche Scientifique - FNRS), under Grants No. 4.4512.10, 1.B.258.20, Belgium;
CECI (Consortium des Equipements de Calcul Intensif), funded by the Fonds de la Recherche Scientifique de Belgique (F.R.S.-FNRS) under Grant No. 2.5020.11 and by the Walloon Region, Belgium;
NSERC (Natural Sciences and Engineering Research Council), funding SAPPJ-2018-0017,  Canada;
MEYS (Ministry of Education, Youth and Sports) funding LM 2018104, Czech Republic;
BMBF (Bundesministerium f\"{u}r Bildung und Forschung) contracts 05H12UM5, 05H15UMCNA and 05H18UMCNA, Germany;
INFN  (Istituto Nazionale di Fisica Nucleare),  Italy;
MIUR (Ministero dell'Istruzione, dell'Universit\`a e della Ricerca),  Italy;
CONACyT  (Consejo Nacional de Ciencia y Tecnolog\'{i}a),  Mexico;
IFA (Institute of Atomic Physics) Romanian 
CERN-RO No. 1/16.03.2016 
and Nucleus Programme PN 19 06 01 04,  Romania;
MESRS  (Ministry of Education, Science, Research and Sport), Slovakia; 
CERN (European Organization for Nuclear Research), Switzerland; 
STFC (Science and Technology Facilities Council), United Kingdom;
NSF (National Science Foundation) Award Numbers 1506088 and 1806430,  U.S.A.;
ERC (European Research Council)  ``UniversaLepto'' advanced grant 268062, ``KaonLepton'' starting grant 336581, Europe.

Individuals have received support from:
Charles University (Research Center UNCE/SCI/013, grant PRIMUS 23/SCI/025), Czech Republic;
Czech Science Foundation (grant 23-06770S);
Ministero dell'Istruzione, dell'Universit\`a e della Ricerca (MIUR  ``Futuro in ricerca 2012''  grant RBFR12JF2Z, Project GAP), Italy;
the Royal Society  (grants UF100308, UF0758946), United Kingdom;
STFC (Rutherford fellowships ST/J00412X/1, ST/M005798/1), United Kingdom;
ERC (grants 268062,  336581 and  starting grant 802836 ``AxScale'');
EU Horizon 2020 (Marie Sk\l{}odowska-Curie grants 701386, 754496, 842407, 893101, 101023808).



\newpage

\newcommand{\orcimg}{\raisebox{-0.3\height}{\includegraphics[height=\fontcharht\font`A]{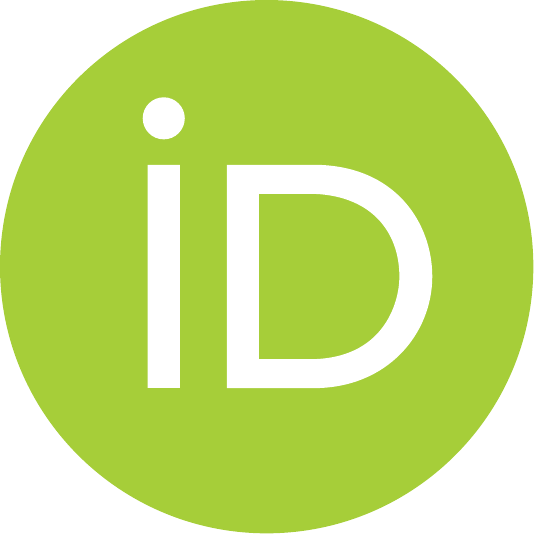}}}
\newcommand{\orcid}[1]{\href{https://orcid.org/#1}{\orcimg}}

\centerline{\bf The NA62 Collaboration} 
\vspace{1.2cm}
%
%

\begin{raggedright}
\noindent
{\bf Universit\'e Catholique de Louvain, Louvain-La-Neuve, Belgium}\\
 E.~Cortina Gil\orcid{0000-0001-9627-699X},
 A.~Kleimenova$\,${\footnotemark[1]}\orcid{0000-0002-9129-4985},
 E.~Minucci$\,${\footnotemark[2]}\orcid{0000-0002-3972-6824},
 S.~Padolski\orcid{0000-0002-6795-7670},
 P.~Petrov, 
 A.~Shaikhiev$\,$\renewcommand{\thefootnote}{\fnsymbol{footnote}}\footnotemark[1]\renewcommand{\thefootnote}{\arabic{footnote}}$^,$$\,${\footnotemark[3]}\orcid{0000-0003-2921-8743},
 R.~Volpe$\,${\footnotemark[4]}\orcid{0000-0003-1782-2978}
\vspace{0.5cm}

{\bf TRIUMF, Vancouver, British Columbia, Canada}\\
 T.~Numao\orcid{0000-0001-5232-6190},
 Y.~Petrov\orcid{0000-0003-2643-8740},
 B.~Velghe\orcid{0000-0002-0797-8381},
 V. W. S.~Wong\orcid{0000-0001-5975-8164}
\vspace{0.5cm}

{\bf University of British Columbia, Vancouver, British Columbia, Canada}\\
 D.~Bryman$\,${\footnotemark[5]}\orcid{0000-0002-9691-0775},
 J.~Fu
\vspace{0.5cm}

{\bf Charles University, Prague, Czech Republic}\\
 Z.~Hives\orcid{0000-0002-5025-993X},
 T.~Husek$\,${\footnotemark[6]}\orcid{0000-0002-7208-9150},
 J.~Jerhot$\,${\footnotemark[7]}\orcid{0000-0002-3236-1471},
 K.~Kampf\orcid{0000-0003-1096-667X},
 M.~Zamkovsky$\,${\footnotemark[8]}\orcid{0000-0002-5067-4789}
\vspace{0.5cm}

{\bf Aix Marseille University, CNRS/IN2P3, CPPM, Marseille, France}\\
 B.~De Martino\orcid{0000-0003-2028-9326},
 M.~Perrin-Terrin\orcid{0000-0002-3568-1956}
\vspace{0.5cm}

{\bf Max-Planck-Institut f\"ur Physik (Werner-Heisenberg-Institut), Garching, Germany}\\
 B.~D\"obrich\orcid{0000-0002-6008-8601},
 S.~Lezki\orcid{0000-0002-6909-774X}
\vspace{0.5cm}

{\bf Institut f\"ur Physik and PRISMA Cluster of Excellence, Universit\"at Mainz, Mainz, Germany}\\
 A. T.~Akmete\orcid{0000-0002-5580-5477},
 R.~Aliberti$\,${\footnotemark[9]}\orcid{0000-0003-3500-4012},
 G.~Khoriauli$\,${\footnotemark[10]}\orcid{0000-0002-6353-8452},
 J.~Kunze,
 D.~Lomidze$\,${\footnotemark[11]}\orcid{0000-0003-3936-6942}, 
 L.~Peruzzo\orcid{0000-0002-4752-6160},
 M.~Vormstein,
 R.~Wanke\orcid{0000-0002-3636-360X}
\vspace{0.5cm}

{\bf Dipartimento di Fisica e Scienze della Terra dell'Universit\`a e INFN, Sezione di Ferrara, Ferrara, Italy}\\
 P.~Dalpiaz,
 M.~Fiorini\orcid{0000-0001-6559-2084},
 A.~Mazzolari\orcid{0000-0003-0804-6778},
 I.~Neri\orcid{0000-0002-9669-1058},
 A.~Norton$\,${\footnotemark[12]}\orcid{0000-0001-5959-5879}, 
 F.~Petrucci\orcid{0000-0002-7220-6919},
 M.~Soldani\orcid{0000-0003-4902-943X},
 H.~Wahl$\,${\footnotemark[13]}\orcid{0000-0003-0354-2465}
\vspace{0.5cm}

{\bf INFN, Sezione di Ferrara, Ferrara, Italy}\\
 L.~Bandiera\orcid{0000-0002-5537-9674},
 A.~Cotta Ramusino\orcid{0000-0003-1727-2478},
 A.~Gianoli\orcid{0000-0002-2456-8667},
 M.~Romagnoni\orcid{0000-0002-2775-6903},
 A.~Sytov\orcid{0000-0001-8789-2440}
\vspace{0.5cm}

{\bf Dipartimento di Fisica e Astronomia dell'Universit\`a e INFN, Sezione di Firenze, Sesto Fiorentino, Italy}\\
 E.~Iacopini\orcid{0000-0002-5605-2497},
 G.~Latino\orcid{0000-0002-4098-3502},
 M.~Lenti\orcid{0000-0002-2765-3955},
 P.~Lo Chiatto\orcid{0000-0002-4177-557X},
 I.~Panichi\orcid{0000-0001-7749-7914},
 A.~Parenti\orcid{0000-0002-6132-5680}
\vspace{0.5cm}

{\bf INFN, Sezione di Firenze, Sesto Fiorentino, Italy}\\
 A.~Bizzeti$\,${\footnotemark[14]}\orcid{0000-0001-5729-5530},
 F.~Bucci\orcid{0000-0003-1726-3838}
\vspace{0.5cm}

{\bf Laboratori Nazionali di Frascati, Frascati, Italy}\\
 A.~Antonelli\orcid{0000-0001-7671-7890},
 G.~Georgiev$\,${\footnotemark[15]}\orcid{0000-0001-6884-3942},
 V.~Kozhuharov$\,${\footnotemark[15]}\orcid{0000-0002-0669-7799},
 G.~Lanfranchi\orcid{0000-0002-9467-8001},
 S.~Martellotti\orcid{0000-0002-4363-7816}, 
 M.~Moulson\orcid{0000-0002-3951-4389},
 T.~Spadaro\orcid{0000-0002-7101-2389},
 G.~Tinti\orcid{0000-0003-1364-844X}
\vspace{0.5cm}

{\bf Dipartimento di Fisica ``Ettore Pancini'' e INFN, Sezione di Napoli, Napoli, Italy}\\
 F.~Ambrosino\orcid{0000-0001-5577-1820},
 T.~Capussela,
 M.~Corvino\orcid{0000-0002-2401-412X},
 M.~D'Errico\orcid{0000-0001-5326-1106},
 D.~Di Filippo\orcid{0000-0003-1567-6786}, 
 R.~Fiorenza$\,${\footnotemark[16]}\orcid{0000-0003-4965-7073},
 R.~Giordano\orcid{0000-0002-5496-7247},
 P.~Massarotti\orcid{0000-0002-9335-9690},
 M.~Mirra\orcid{0000-0002-1190-2961},
 M.~Napolitano\orcid{0000-0003-1074-9552}, 
 I.~Rosa\orcid{0009-0002-7564-1825},
 G.~Saracino\orcid{0000-0002-0714-5777}
\vspace{0.5cm}

{\bf Dipartimento di Fisica e Geologia dell'Universit\`a e INFN, Sezione di Perugia, Perugia, Italy}\\
 G.~Anzivino\orcid{0000-0002-5967-0952},
 F.~Brizioli$\,${\footnotemark[8]}\orcid{0000-0002-2047-441X},
 E.~Imbergamo,
 R.~Lollini\orcid{0000-0003-3898-7464},
 R.~Piandani$\,${\footnotemark[17]}\orcid{0000-0003-2226-8924},
 C.~Santoni\orcid{0000-0001-7023-7116}
\vspace{0.5cm}

{\bf INFN, Sezione di Perugia, Perugia, Italy}\\
 M.~Barbanera\orcid{0000-0002-3616-3341},
 P.~Cenci\orcid{0000-0001-6149-2676},
 B.~Checcucci\orcid{0000-0002-6464-1099},
 P.~Lubrano\orcid{0000-0003-0221-4806},
 M.~Lupi$\,${\footnotemark[18]}\orcid{0000-0001-9770-6197}, 
 M.~Pepe\orcid{0000-0001-5624-4010},
 M.~Piccini\orcid{0000-0001-8659-4409}
\vspace{0.5cm}

{\bf Dipartimento di Fisica dell'Universit\`a e INFN, Sezione di Pisa, Pisa, Italy}\\
 F.~Costantini\orcid{0000-0002-2974-0067},
 L.~Di Lella$\,${\footnotemark[13]}\orcid{0000-0003-3697-1098},
 N.~Doble$\,${\footnotemark[13]}\orcid{0000-0002-0174-5608},
 M.~Giorgi\orcid{0000-0001-9571-6260},
 S.~Giudici\orcid{0000-0003-3423-7981}, 
 G.~Lamanna\orcid{0000-0001-7452-8498},
 E.~Lari\orcid{0000-0003-3303-0524},
 E.~Pedreschi\orcid{0000-0001-7631-3933},
 M.~Sozzi\orcid{0000-0002-2923-1465}
\vspace{0.5cm}

{\bf INFN, Sezione di Pisa, Pisa, Italy}\\
 C.~Cerri,
 R.~Fantechi\orcid{0000-0002-6243-5726},
 L.~Pontisso$\,${\footnotemark[19]}\orcid{0000-0001-7137-5254},
 F.~Spinella\orcid{0000-0002-9607-7920}
\vspace{0.5cm}

{\bf Scuola Normale Superiore e INFN, Sezione di Pisa, Pisa, Italy}\\
 I.~Mannelli\orcid{0000-0003-0445-7422}
\vspace{0.5cm}

{\bf Dipartimento di Fisica, Sapienza Universit\`a di Roma e INFN, Sezione di Roma I, Roma, Italy}\\
 G.~D'Agostini\orcid{0000-0002-6245-875X},
 M.~Raggi\orcid{0000-0002-7448-9481}
\vspace{0.5cm}

{\bf INFN, Sezione di Roma I, Roma, Italy}\\
 A.~Biagioni\orcid{0000-0001-5820-1209},
 P.~Cretaro\orcid{0000-0002-2229-149X},
 O.~Frezza\orcid{0000-0001-8277-1877},
 E.~Leonardi\orcid{0000-0001-8728-7582},
 A.~Lonardo\orcid{0000-0002-5909-6508}, 
 M.~Turisini\orcid{0000-0002-5422-1891},
 P.~Valente\orcid{0000-0002-5413-0068},
 P.~Vicini\orcid{0000-0002-4379-4563}
\vspace{0.5cm}

{\bf INFN, Sezione di Roma Tor Vergata, Roma, Italy}\\
 R.~Ammendola\orcid{0000-0003-4501-3289},
 V.~Bonaiuto$\,${\footnotemark[20]}\orcid{0000-0002-2328-4793},
 A.~Fucci,
 A.~Salamon\orcid{0000-0002-8438-8983},
 F.~Sargeni$\,${\footnotemark[21]}\orcid{0000-0002-0131-236X}
\vspace{0.5cm}

{\bf Dipartimento di Fisica dell'Universit\`a e INFN, Sezione di Torino, Torino, Italy}\\
 R.~Arcidiacono$\,${\footnotemark[22]}\orcid{0000-0001-5904-142X},
 B.~Bloch-Devaux\orcid{0000-0002-2463-1232},
 M.~Boretto$\,${\footnotemark[8]}\orcid{0000-0001-5012-4480},
 E.~Menichetti\orcid{0000-0001-7143-8200},
 E.~Migliore\orcid{0000-0002-2271-5192},
 D.~Soldi\orcid{0000-0001-9059-4831}
\vspace{0.5cm}

{\bf INFN, Sezione di Torino, Torino, Italy}\\
 C.~Biino\orcid{0000-0002-1397-7246},
 A.~Filippi\orcid{0000-0003-4715-8748},
 F.~Marchetto\orcid{0000-0002-5623-8494}
\vspace{0.5cm}

{\bf Instituto de F\'isica, Universidad Aut\'onoma de San Luis Potos\'i, San Luis Potos\'i, Mexico}\\
 A.~Briano Olvera\orcid{0000-0001-6121-3905},
 J.~Engelfried\orcid{0000-0001-5478-0602},
 N.~Estrada-Tristan$\,${\footnotemark[23]}\orcid{0000-0003-2977-9380},
 M. A.~Reyes Santos$\,${\footnotemark[23]}\orcid{0000-0003-1347-2579},
 K. A.~Rodriguez Rivera\orcid{0000-0001-5723-9176}
\vspace{0.5cm}

{\bf Horia Hulubei National Institute for R\&D in Physics and Nuclear Engineering, Bucharest-Magurele, Romania}\\
 P.~Boboc\orcid{0000-0001-5532-4887},
 A. M.~Bragadireanu,
 S. A.~Ghinescu\orcid{0000-0003-3716-9857},
 O. E.~Hutanu
\vspace{0.5cm}

{\bf Faculty of Mathematics, Physics and Informatics, Comenius University, Bratislava, Slovakia}\\
 L.~Bician$\,${\footnotemark[24]}\orcid{0000-0001-9318-0116},
 T.~Blazek\orcid{0000-0002-2645-0283},
 V.~Cerny\orcid{0000-0003-1998-3441},
 Z.~Kucerova$\,${\footnotemark[8]}\orcid{0000-0001-8906-3902}
\vspace{0.5cm}

\vspace{0.5cm}
{\bf CERN, European Organization for Nuclear Research, Geneva, Switzerland}\\
 J.~Bernhard\orcid{0000-0001-9256-971X},
 A.~Ceccucci\orcid{0000-0002-9506-866X},
 M.~Ceoletta\orcid{0000-0002-2532-0217},
 H.~Danielsson\orcid{0000-0002-1016-5576},
 N.~De Simone$\,${\footnotemark[25]}, 
 F.~Duval,
 L.~Federici\orcid{0000-0002-3401-9522},
 E.~Gamberini\orcid{0000-0002-6040-4985},
 L.~Gatignon$\,${\footnotemark[3]}\orcid{0000-0001-6439-2945},
 R.~Guida, 
 F.~Hahn$\,$\renewcommand{\thefootnote}{\fnsymbol{footnote}}\footnotemark[2]\renewcommand{\thefootnote}{\arabic{footnote}},
 E.~B.~Holzer\orcid{0000-0003-2622-6844},
 B.~Jenninger,
 M.~Koval$\,${\footnotemark[24]}\orcid{0000-0002-6027-317X},
 P.~Laycock$\,${\footnotemark[26]}\orcid{0000-0002-8572-5339}, 
 G.~Lehmann Miotto\orcid{0000-0001-9045-7853},
 P.~Lichard\orcid{0000-0003-2223-9373},
 A.~Mapelli\orcid{0000-0002-4128-1019},
 R.~Marchevski$\,${\footnotemark[1]}\orcid{0000-0003-3410-0918},
 K.~Massri\orcid{0000-0001-7533-6295}, 
 M.~Noy,
 V.~Palladino\orcid{0000-0002-9786-9620},
 J.~Pinzino$\,${\footnotemark[27]}\orcid{0000-0002-7418-0636},
 V.~Ryjov,
 S.~Schuchmann\orcid{0000-0002-8088-4226}, 
 S.~Venditti
\vspace{0.5cm}

{\bf School of Physics and Astronomy, University of Birmingham, Birmingham, United Kingdom}\\
 T.~Bache\orcid{0000-0003-4520-830X},
 M. B.~Brunetti$\,${\footnotemark[28]}\orcid{0000-0003-1639-3577},
 V.~Duk$\,${\footnotemark[4]}\orcid{0000-0001-6440-0087},
 V.~Fascianelli$\,${\footnotemark[29]},
 J. R.~Fry\orcid{0000-0002-3680-361X}, 
 F.~Gonnella\orcid{0000-0003-0885-1654},
 E.~Goudzovski\orcid{0000-0001-9398-4237},
 J.~Henshaw\orcid{0000-0001-7059-421X},
 L.~Iacobuzio,
 C.~Kenworthy\orcid{0009-0002-8815-0048}, 
 C.~Lazzeroni\orcid{0000-0003-4074-4787},
 N.~Lurkin$\,${\footnotemark[30]}\orcid{0000-0002-9440-5927},
 F.~Newson,
 C.~Parkinson\orcid{0000-0003-0344-7361},
 A.~Romano\orcid{0000-0003-1779-9122}, 
 J.~Sanders\orcid{0000-0003-1014-094X},
 A.~Sergi$\,${\footnotemark[31]}\orcid{0000-0001-9495-6115},
 A.~Sturgess\orcid{0000-0002-8104-5571},
 J.~Swallow$\,${\footnotemark[8]}\orcid{0000-0002-1521-0911},
 A.~Tomczak\orcid{0000-0001-5635-3567}
\vspace{0.5cm}

{\bf School of Physics, University of Bristol, Bristol, United Kingdom}\\
 H.~Heath\orcid{0000-0001-6576-9740},
 R.~Page,
 S.~Trilov\orcid{0000-0003-0267-6402}
\vspace{0.5cm}

{\bf School of Physics and Astronomy, University of Glasgow, Glasgow, United Kingdom}\\
 B.~Angelucci,
 D.~Britton\orcid{0000-0001-9998-4342},
 C.~Graham\orcid{0000-0001-9121-460X},
 D.~Protopopescu\orcid{0000-0002-8047-6513}
\vspace{0.5cm}

{\bf Physics Department, University of Lancaster, Lancaster, United Kingdom}\\
 J.~Carmignani$\,${\footnotemark[32]}\orcid{0000-0002-1705-1061},
 J. B.~Dainton,
 R. W. L.~Jones\orcid{0000-0002-6427-3513},
 G.~Ruggiero$\,${\footnotemark[33]}\orcid{0000-0001-6605-4739}
\vspace{0.5cm}

{\bf School of Physical Sciences, University of Liverpool, Liverpool, United Kingdom}\\
 L.~Fulton,
 D.~Hutchcroft\orcid{0000-0002-4174-6509},
 E.~Maurice$\,${\footnotemark[34]}\orcid{0000-0002-7366-4364},
 B.~Wrona\orcid{0000-0002-1555-0262}
\vspace{0.5cm}

{\bf Physics and Astronomy Department, George Mason University, Fairfax, Virginia, USA}\\
 A.~Conovaloff,
 P.~Cooper,
 D.~Coward$\,${\footnotemark[35]}\orcid{0000-0001-7588-1779},
 P.~Rubin\orcid{0000-0001-6678-4985}
\vspace{0.5cm}

{\bf Authors affiliated with an Institute or an international laboratory covered by a cooperation agreement with CERN}\\
 A.~Baeva,
 D.~Baigarashev$\,${\footnotemark[36]}\orcid{0000-0001-6101-317X},
 D.~Emelyanov,
 T.~Enik\orcid{0000-0002-2761-9730},
 V.~Falaleev$\,${\footnotemark[4]}\orcid{0000-0003-3150-2196}, 
 S.~Fedotov,
 K.~Gorshanov\orcid{0000-0001-7912-5962},
 E.~Gushchin\orcid{0000-0001-8857-1665},
 V.~Kekelidze\orcid{0000-0001-8122-5065},
 D.~Kereibay, 
 S.~Kholodenko$\,${\footnotemark[27]}\orcid{0000-0002-0260-6570},
 A.~Khotyantsev,
 A.~Korotkova,
 Y.~Kudenko\orcid{0000-0003-3204-9426},
 V.~Kurochka, 
 V.~Kurshetsov\orcid{0000-0003-0174-7336},
 L.~Litov$\,${\footnotemark[15]}\orcid{0000-0002-8511-6883},
 D.~Madigozhin\orcid{0000-0001-8524-3455},
 M.~Medvedeva,
 A.~Mefodev, 
 M.~Misheva$\,${\footnotemark[37]},
 N.~Molokanova,
 S.~Movchan,
 V.~Obraztsov\orcid{0000-0002-0994-3641},
 A.~Okhotnikov\orcid{0000-0003-1404-3522}, 
 A.~Ostankov$\,$\renewcommand{\thefootnote}{\fnsymbol{footnote}}\footnotemark[2]\renewcommand{\thefootnote}{\arabic{footnote}},
 I.~Polenkevich,
 Yu.~Potrebenikov\orcid{0000-0003-1437-4129},
 A.~Sadovskiy\orcid{0000-0002-4448-6845},
 V.~Semenov$\,$\renewcommand{\thefootnote}{\fnsymbol{footnote}}\footnotemark[2]\renewcommand{\thefootnote}{\arabic{footnote}}, 
 S.~Shkarovskiy,
 V.~Sugonyaev\orcid{0000-0003-4449-9993},
 O.~Yushchenko\orcid{0000-0003-4236-5115},
 A.~Zinchenko$\,$\renewcommand{\thefootnote}{\fnsymbol{footnote}}\footnotemark[2]\renewcommand{\thefootnote}{\arabic{footnote}}
\vspace{0.5cm}

\end{raggedright}

%
%

\setcounter{footnote}{0}
\newlength{\basefootnotesep}
\setlength{\basefootnotesep}{\footnotesep}

\renewcommand{\thefootnote}{\fnsymbol{footnote}}
\noindent
$^{\footnotemark[1]}${Corresponding author:  A.~Shaikhiev, email: artur.shaikhiev@cern.ch}\\
$^{\footnotemark[2]}${Deceased}\\
\renewcommand{\thefootnote}{\arabic{footnote}}
$^{1}${Present address: Ecole Polytechnique F\'ed\'erale Lausanne, CH-1015 Lausanne, Switzerland} \\
$^{2}${Present address: Syracuse University, Syracuse, NY 13244, USA} \\
$^{3}${Present address: Physics Department, University of Lancaster, Lancaster,  LA1 4YB, UK} \\
$^{4}${Present address: INFN, Sezione di Perugia, I-06100 Perugia, Italy} \\
$^{5}${Also at TRIUMF, Vancouver, British Columbia, V6T 2A3, Canada} \\
$^{6}${Also at School of Physics and Astronomy, University of Birmingham, Birmingham, B15 2TT, UK} \\
$^{7}${Present address: Max-Planck-Institut f\"ur Physik (Werner-Heisenberg-Institut), Garching,  \\
D-85748, Germany} \\
$^{8}${Present address: CERN, European Organization for Nuclear Research, CH-1211 Geneva 23, Switzerland} \\
$^{9}${Present address: Institut f\"ur Kernphysik and Helmholtz Institute Mainz, Universit\"at Mainz, Mainz, D-55099, Germany} \\
$^{10}${Present address: Universit\"at W\"urzburg, D-97070 W\"urzburg, Germany} \\
$^{11}${Present address: European XFEL GmbH, D-22869 Schenefeld, Germany} \\
$^{12}${Present address: School of Physics and Astronomy, University of Glasgow, Glasgow, G12 8QQ, UK} \\
$^{13}${Present address: Institut f\"ur Physik and PRISMA Cluster of Excellence, Universit\"at Mainz, D-55099 Mainz, Germany} \\
$^{14}${Also at Dipartimento di Scienze Fisiche, Informatiche e Matematiche, Universit\`a di Modena e Reggio Emilia, I-41125 Modena, Italy} \\
$^{15}${Also at Faculty of Physics, University of Sofia, BG-1164 Sofia, Bulgaria} \\
$^{16}${Present address: Scuola Superiore Meridionale e INFN, Sezione di Napoli, I-80138 Napoli, Italy} \\
$^{17}${Present address: Instituto de F\'isica, Universidad Aut\'onoma de San Luis Potos\'i, 78240 San Luis Potos\'i, Mexico} \\
$^{18}${Present address: Institut am Fachbereich Informatik und Mathematik, Goethe Universit\"at, D-60323 Frankfurt am Main, Germany} \\
$^{19}${Present address: INFN, Sezione di Roma I, I-00185 Roma, Italy} \\
$^{20}${Also at Department of Industrial Engineering, University of Roma Tor Vergata, I-00173 Roma, Italy} \\
$^{21}${Also at Department of Electronic Engineering, University of Roma Tor Vergata, I-00173 Roma, Italy} \\
$^{22}${Also at Universit\`a degli Studi del Piemonte Orientale, I-13100 Vercelli, Italy} \\
$^{23}${Also at Universidad de Guanajuato, 36000 Guanajuato, Mexico} \\
$^{24}${Present address: Charles University, 116 36 Prague 1, Czech Republic} \\
$^{25}${Present address: DESY, D-15738 Zeuthen, Germany} \\
$^{26}${Present address: Brookhaven National Laboratory, Upton, NY 11973, USA} \\
$^{27}${Present address: INFN, Sezione di Pisa, I-56100 Pisa, Italy} \\
$^{28}${Present address: Department of Physics, University of Warwick, Coventry, CV4 7AL, UK} \\
$^{29}${Present address: Center for theoretical neuroscience, Columbia University, New York,  \\
NY 10027, USA} \\
$^{30}${Present address: Universit\'e Catholique de Louvain, B-1348 Louvain-La-Neuve, Belgium} \\
$^{31}${Present address: Dipartimento di Fisica dell'Universit\`a e INFN, Sezione di Genova, I-16146 Genova, Italy} \\
$^{32}${Present address: School of Physical Sciences, University of Liverpool, Liverpool, L69 7ZE, UK} \\
$^{33}${Present address: Dipartimento di Fisica e Astronomia dell'Universit\`a e INFN, Sezione di Firenze, I-50019 Sesto Fiorentino, Italy} \\
$^{34}${Present address: Laboratoire Leprince Ringuet, F-91120 Palaiseau, France} \\
$^{35}${Also at SLAC National Accelerator Laboratory, Stanford University, Menlo Park, CA 94025, USA} \\
$^{36}${Also at L. N. Gumilyov Eurasian National University, 010000 Nur-Sultan, Kazakhstan} \\
$^{37}${Present address: Institute of Nuclear Research and Nuclear Energy of Bulgarian Academy of Science (INRNE-BAS), BG-1784 Sofia, Bulgaria} \\

\end{document}